\documentclass[a4paper,11pt]{article}
\pdfoutput=1
\usepackage{jcappub}
\usepackage{lineno}
\usepackage{orcidlink}
\usepackage[section]{placeins}
\usepackage{float}


\title{Cosmic-web quenching with DESI DR1: T-Web environments and mass-dependent red/blue classification}
\author[a]{Hafiz Inam Ullah\orcidlink{0009-0005-5796-6054}}
\author[a]{,Muhammad Awais\orcidlink{0009-0000-9174-8492}}
\author[b]{,Tonatiuh Matos}
\author[c,1]{,\note{Corresponding author}John F. Su\'arez-P\'erez\orcidlink{0000-0002-0896-8134}%
}

\affiliation[a]{Escuela de F\'isica y Matem\'aticas del Instituto Polit\'ecnico Nacional, M\'exico}

\affiliation[b]{Departamento de F\'isica, Centro de Investigaci\'on y de Estudios Avanzados del Instituto Polit\'ecnico Nacional, M\'exico}

\affiliation[c]{Tecnologico de Monterrey, Escuela de Ingenier\'ia y Ciencias, Zapopan, M\'exico}

\emailAdd{hinamu2500@alumno.ipn.mx}
\emailAdd{mawais2500@alumno.ipn.mx}
\emailAdd{tonatiuh.matos@cinvestav.mx}
\emailAdd{jf.suarez@tec.mx}

\abstract{
We study DESI DR1 galaxies to quantify colour dependence on cosmic web environment for three tracers spanning complementary regimes: BGS ($0.15\le z<0.55$), LRG ($0.6\le z<0.9$), and ELG ($0.6\le z<1.6$). Web environments are reconstructed with the tidal-tensor (T-Web) formalism on a $256^3$ grid in an $800\,Mpc$ cube and classified into voids, sheets, filaments, and knots. Sheets and filaments dominate volume ($\sim 45$--$48\%$ and $\sim 37$--$40\%$), voids $\sim 6$--$16\%$ knots $\sim 4$--$6\%$. A mass-dependent Otsu method separates red and blue populations. The BGS red fraction evolves non-monotonically: at $z\approx0.20$, voids ($13.89\pm5.76\%$), sheets ($6.13\pm1.27\%$), filaments ($9.24\pm1.66\%$), knots ($6.12\pm3.42\%$); at $z\approx0.30$, values range from $0.63\pm0.44\%$ to $2.01\pm0.99\%$; at $z\approx0.50$, from $17.93\pm0.44\%$ to $19.63\pm1.08\%$; environmental differences are small. LRGs show environment-dependent quenching: at $z\approx0.66$, knots ($65.90\pm0.45\%$), voids ($62.40\pm1.81\%$), filaments ($60.21\pm0.48\%$), sheets ($58.37\pm3.15\%$); by $z\approx0.88$, these converge to $\sim 68$--$70\%$. ELGs exhibit strong redshift evolution: filaments drop from $55.18\pm0.31\%$ at $z\approx0.65$ to $33.22\pm0.21\%$ at $z\approx0.95$; voids and sheets show similar declines, with weak and non-monotonic. High-mass selection increases red fractions but preserves trends. Relative red and blue fractions (RRF/RBF) show filaments and sheets host the largest shares of both red and blue galaxies; knots contribute less despite elevated red fractions. The $(g-r)$ colour distributions reveal an enhanced red component in knots and bluer colours in voids, with the clearest bimodality at low redshift. Overall, stellar mass drives the primary quenching trend, while environment provides a systematic secondary modulation, strongest in dense knots and at lower stellar masses.
}

\begin{document}
\maketitle
\flushbottom

\section{Introduction}
The large scale structure of the Universe, shaped by gravitational instability acting on primordial density fluctuations, forms a connected network of overdense and underdense regions known as the cosmic web \citep{Bond1996, ZELDOVICH1970}. 
This web is commonly described in terms of four morphological environments voids, sheets, filaments, and knots each characterised by distinct density contrasts, dynamical flows, and tidal field signatures \citep{Hahn2007, Forero-Romero2009, Libeskind2018}. 
Because galaxies form and evolve within this scaffold, their observable properties reflect both internal processes and the influence of their surrounding environment across a wide range of scales \citep{White1987, Kauffmann2004}. 
A central goal of galaxy formation theory is therefore to quantify how large scale environment modulates star formation and quiescence, and how this modulation evolves with cosmic time.
Previous studies have established that both stellar mass and environment play important roles in galaxy quenching, with mass being the primary driver and environment providing a secondary modulation \citep{Peng2010, Contini2020}.

The Dark Energy Spectroscopic Instrument (DESI) enables a new class of precision measurements in both cosmology and galaxy evolution. 
Its first public data release, DESI DR1, provides more than 13 million extragalactic redshifts nearly four times the number of unique extragalactic spectra released by all Sloan Digital Sky Survey programs combined \citep{DESICollaboration2025}. 
With contiguous sky coverage of $\sim 10\,000~\mathrm{deg}^2$ and galaxy redshifts extending to $z\lesssim 1.6$, DESI DR1 offers an unprecedented opportunity to map the cosmic web with high statistical power and to test simulation-driven predictions for environmental trends using large observational samples.

A wide range of algorithms have been developed to classify the cosmic web.
These include geometric methods such as the \textsc{DisPerSE} skeleton \citep{Sousbie2010}, dynamical approaches based on the tidal or deformation tensor \citep{Forero-Romero2009, Hoffman2012}, and more recent graph- and machine-learning-based methods that exploit the topology of galaxy distributions, such as the $\beta$-skeleton approach introduced by \cite{Suarez-Perez2021} and the ASTRA algorithm \citep{Forero-Romero2025}, which uses random points to trace underdense regions and robustly identify cosmic voids.
Among these, the tidal-tensor (T-Web) formalism provides a physically motivated, scale-dependent description of the local gravitational field: 
the number of collapsing directions inferred from the tidal eigenvalues naturally partitions the density field into voids, sheets, filaments, and knots. 
Its computational efficiency and direct dynamical interpretability make it well suited for application to modern, large-volume spectroscopic surveys and for connecting galaxy properties to the anisotropic character of gravitational collapse.

Recent hydrodynamical simulations, particularly the IllustrisTNG project \citep{Nelson2019}, have advanced quantitative predictions for how the cosmic web modulates galaxy quenching \citep{Pandey2025}.
Using a mass-dependent colour classification based on Otsu's method, \citep{Otsu1979, Pandey2025} argued that environmental quenching is most efficient below a characteristic stellar mass $\log(M_\ast/M_\odot)\approx 10.5$, while above this threshold internal processes (e.g.\ active galactic nucleus feedback) dominate. 
They further introduced relative red/blue fractions (RRF/RBF) to quantify how quenched and star-forming populations are partitioned across the web, and reported a decline in environmental differentiation with increasing redshift.
A key open question is whether these patterns, derived in simulations under specific modelling assumptions, are realised in observational samples with comparable statistical power and dynamic range.

In this paper we perform a systematic observational test of these predictions using DESI DR1.
We employ three complementary galaxy samples Luminous Red Galaxies (LRGs), Emission-Line Galaxies (ELGs), and the Bright Galaxy Survey (BGS) which span different stellar-mass regimes, bias factors, and redshift intervals. 
We reconstruct the cosmic web for each tracer using the T-Web formalism and investigate the scale dependence of the resulting environment classification. 
We then classify galaxies into red and blue populations using a mass-dependent implementation of Otsu's method in the colour mass plane, applied independently in each redshift bin, and quantify environmental trends through RF/BF and RRF/RBF as functions of redshift and stellar mass.

The paper is organised as follows. 
Section~\ref{sec:data} describes the DESI DR1 galaxy samples, the construction of the density field, and the value-added catalogues used for stellar masses and colours.
Section~\ref{sec:method} details the T-Web environment classification and the Otsu-based red/blue separation. 
In Section~\ref{sec:results} we present the cosmic web maps, the scale-dependence analysis, the colour--mass diagrams with Otsu thresholds, and the evolution of red/blue fractions across environments.
Finally, Section~\ref{sec:conclusions} summarises our conclusions and outlines future prospects with upcoming DESI data releases.

\section{Data}\label{sec:data}

\subsection{DESI DR1 galaxy samples}
\label{sec:data_samples}

We use spectroscopic galaxy samples from the first public data release of the Dark Energy Spectroscopic Instrument \citep{DESI_EDR_2023}, selected from the standard DESI clustering catalogs \citep{DESI_2024_II}. 
Our analysis employs three tracer populations: luminous red galaxies (LRGs), emission-line galaxies (ELGs), and the Bright Galaxy Survey (BGS).
These tracers span different redshift ranges, galaxy types, and large-scale bias, enabling a comparative study of galaxy
evolution across cosmic-web environments.

For each tracer, we use two types of catalogs provided by the DESI clustering pipeline \citep{DESI_2024_II}: 
(i) the data catalog, containing the observed galaxies,
and 
(ii) the corresponding random catalogs, containing unclustered points that follow the same survey angular footprint, imaging mask, and radial selection function as the data catalog, containing the observed galaxies.

Concretely, for the BGS tracer we use the file \texttt{{BGS\_ANY\_NGC\_clustering.dat.fits}} as the
galaxy catalog and the set of files \texttt{BGS\_ANY\_NGC\{0..17\}clustering.ran.fits} as the random catalogs. 
Analogous \texttt{clustering.dat.fits} and \texttt{clustering.ran.fits} files are used for the LRG and ELG tracers.
The multiple random files are provided in chunks for practical file-size reasons and are concatenated in our analysis. 
Throughout this work we restrict to the NGC (North Galactic Cap) footprint to maximize contiguity and simplify survey-window handling.

Each catalog provides sky positions (Right Ascension ($\alpha$) and Declination ($\delta$)) and spectroscopic redshifts (z).
We also use the standard DESI clustering weights supplied for large-scale structure analyses (e.g. completeness and imaging systematics weights; \citep{Ross2020, Burden2024}).
These weights are included consistently in the number-density estimates used to build overdensity fields and in the galaxy counts used for red/blue fraction measurements.

The redshift ranges adopted in this work are: 
BGS: $0.15 \le z < 0.55$ \citep{Hahn2023};
LRG: $0.6 \le z < 0.9$ \citep{Zhou2023};
ELG: $0.6 \le z \le 1.6$ \citep{Raichoor2023}.
These choices follow the redshift intervals where each tracer is well populated and where the colour–mass classification remains stable \cite{DESI_DR1_2025}.

\subsection{Stellar masses and colours}
\label{sec:data_mstar_colour}

Stellar masses and additional spectral/photometric information are taken from the \\\texttt{DESI\_DR1\_value-added\_product\_dr1\_galaxy\_stellarmass\_lineinfo\_v1.0.fits}, which we cross-match to the clustering catalogues using TARGETID identifier \citep{DESI_DR1_2025,DESI_MassEMLines_VAC}. 
We use $\log(M_\ast/M_\odot)$ as our stellar-mass estimate and the observed colour (g-r) as the primary colour for the red/blue classification.
We adopt (g-r) because it is uniformly available for all three tracers in the DR1 products used here; the red/blue divider is recomputed independently in each redshift bin to account for colour evolution.

\subsection{Comoving coordinates and fiducial cosmology}
\label{sec:data_coords}

To construct three-dimensional density fields for the T-Web reconstruction, 
we convert each galaxy position $(\alpha,\delta,z)$ into comoving Cartesian coordinates using a fiducial flat $\Lambda$CDM cosmology consistent with Planck 2018. 
We adopt $H_0 = 67.4~\mathrm{km\,s^{-1}\,Mpc^{-1}}$, $\Omega_m=0.315$, and $\Omega_\Lambda=0.685$.
The comoving radial distance is $\chi(z)$, and the Cartesian coordinates are
\begin{equation}
x=\chi(z)\cos\delta\cos\alpha,\qquad
y=\chi(z)\cos\delta\sin\alpha,\qquad
z=\chi(z)\sin\delta,
\end{equation}
where $\alpha$ and $\delta$ are right ascension and declination, respectively.
All distances are expressed in comoving Mpc throughout.

\section{Method}\label{sec:method}

The primary objective of this study is to classify galaxies in the DESI DR1 public spectroscopic sample into red and blue populations as functions of stellar mass and redshift, and to quantify how their colour evolution depends on the large-scale cosmic-web environment. Using the three main DESI tracers the Bright Galaxy Survey (BGS), luminous red galaxies (LRG), and emission line galaxies (ELG) we measure the red fraction (RF) and blue fraction (BF) in four T-Web environments (voids, sheets, filaments, and knots) across multiple redshift intervals and stellar mass bins. In addition, we compute the probability density functions (PDFs) of the $(g-r)$ colour distribution in each environment and redshift bin, as well as the relative red and blue fractions, RRF and RBF, in order to quantify how the red and blue populations are distributed across the cosmic web.

\subsection{Identifying different morphological environments of the cosmic web with T-Web}
\label{sec:tweb_method_general}

We classify galaxies into morphological environments using the tidal-tensor (T-Web) formalism \citep{Hahn2007, Forero-Romero2009}. 
In this approach, the local web morphology is determined by the eigenvalues of the deformation tensor (the Hessian of the gravitational potential) computed from the smoothed overdensity field. 
Physically, the method quantifies the anisotropic collapse/expansion of matter along the principal axes of the tidal field, enabling a dynamical classification into voids, sheets, filaments, and knots (clusters) \citep{Hahn2007, Forero-Romero2009}.

\paragraph{Density and overdensity field.}
For each DESI tracer sample, we construct a three-dimensional density field on a Cartesian grid using
the Cloud-In-Cell (CIC) mass-assignment scheme \citep{HockneyEastwood1981}.
To account for the survey selection function (angular mask and radial completeness), we build the corresponding CIC field from the random catalogues on the same grid. 
Denoting the weighted CIC number-density fields by $n_g(\mathbf{x})$ (galaxies) and $n_r(\mathbf{x})$ (random), we estimate the overdensity as 
\begin{equation}
\varrho(\mathbf{x}) \equiv \frac{n_g(\mathbf{x})}{\gamma\,n_r(\mathbf{x})}-1,
\qquad
\gamma \equiv \frac{\sum w_g}{\sum w_r},
\label{eq:delta_randoms}
\end{equation}
where $\gamma$ normalizes the randoms to the data and $w_g$ and $w_r$ are the total weights applied to galaxies and randoms, respectively. 
In constructing $n_g(\mathbf{x})$ and $n_r(\mathbf{x})$, we include the standard DESI clustering weights to account for observational systematics and completeness \citep{Ross2020,Burden2024}.
The resulting overdensity field is smoothed with an isotropic Gaussian filter of width $R_s$ to suppress shot noise and to emphasize the quasi-linear cosmic-web morphology.

\paragraph{Gravitational potential.}
The gravitational potential $\Phi(\mathbf{x})$ is obtained by solving the Poisson equation
\begin{equation}
\nabla^2 \Phi(\mathbf{x}) = \varrho(\mathbf{x}).
\label{eq:poisson}
\end{equation}
We solve Eq.~\eqref{eq:poisson} in Fourier space, where it becomes algebraic:
\begin{equation}
\hat{\Phi}(\mathbf{k}) = -\frac{\hat{\varrho}(\mathbf{k})}{k^2},
\qquad (k\neq 0),
\label{eq:phi_fourier}
\end{equation}
with hats denoting Fourier transforms and $k \equiv |\mathbf{k}|$. 
The $k=0$ mode is set to zero, corresponding to an arbitrary additive constant in $\Phi$. 
To mitigate boundary effects from the survey window, we restrict environment assignment to grid cells that are well sampled by the random catalogue (i.e.\ with non-negligible $n_r$).

\paragraph{Deformation (tidal) tensor.}
The T-Web deformation tensor is defined as the Hessian of the potential,
\begin{equation}
T_{ij}(\mathbf{x}) \equiv \frac{\partial^2 \Phi(\mathbf{x})}{\partial x_i\,\partial x_j},
\label{eq:tij}
\end{equation}
where $x_i$ and $x_j$ are Cartesian coordinates. 
We evaluate $T_{ij}$ on the grid (equivalently by using Fourier-space derivatives) and diagonalize it to obtain the ordered eigenvalues $\lambda_1 > \lambda_2 > \lambda_3$ in each grid cell.

\paragraph{Cosmic-web classification.}
Following the standard T-Web prescription, the environment is determined by the number of eigenvalues
exceeding a threshold $\lambda_{\rm th}$ \citep{Forero-Romero2009}. 
In this work we adopt $\lambda_{\rm th}=0$, such that
\begin{itemize}
\item \textbf{Void:} $\lambda_1,\lambda_2,\lambda_3 < 0$,
\item \textbf{Sheet:} $\lambda_1 > 0$, $\lambda_2,\lambda_3 < 0$,
\item \textbf{Filament:} $\lambda_1,\lambda_2 > 0$, $\lambda_3 < 0$,
\item \textbf{Knot (cluster):} $\lambda_1,\lambda_2,\lambda_3 > 0$.
\end{itemize}
Finally, each galaxy is assigned a web environment by mapping its comoving position to the grid and adopting the environment label of the corresponding cell in the environment map.

The specific parameters adopted for each tracer (box size, grid resolution, smoothing scales) are described in the following subsections.

\subsection{Tidal Web parameters}
\label{subsec:tweb}

\begin{figure}[!htbp]
\centering
\includegraphics[width=0.7\textwidth]{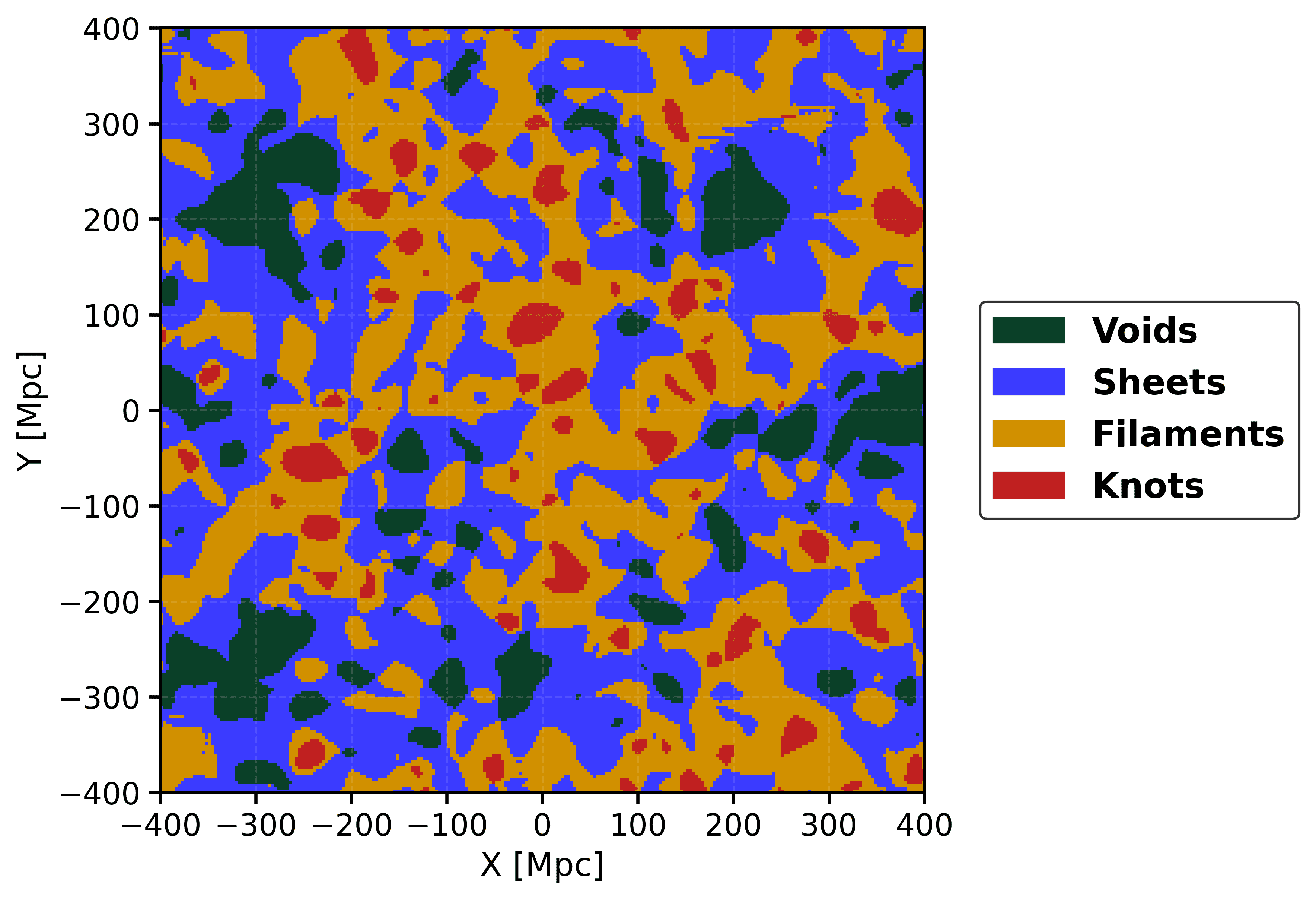}
\includegraphics[width=0.7\textwidth]{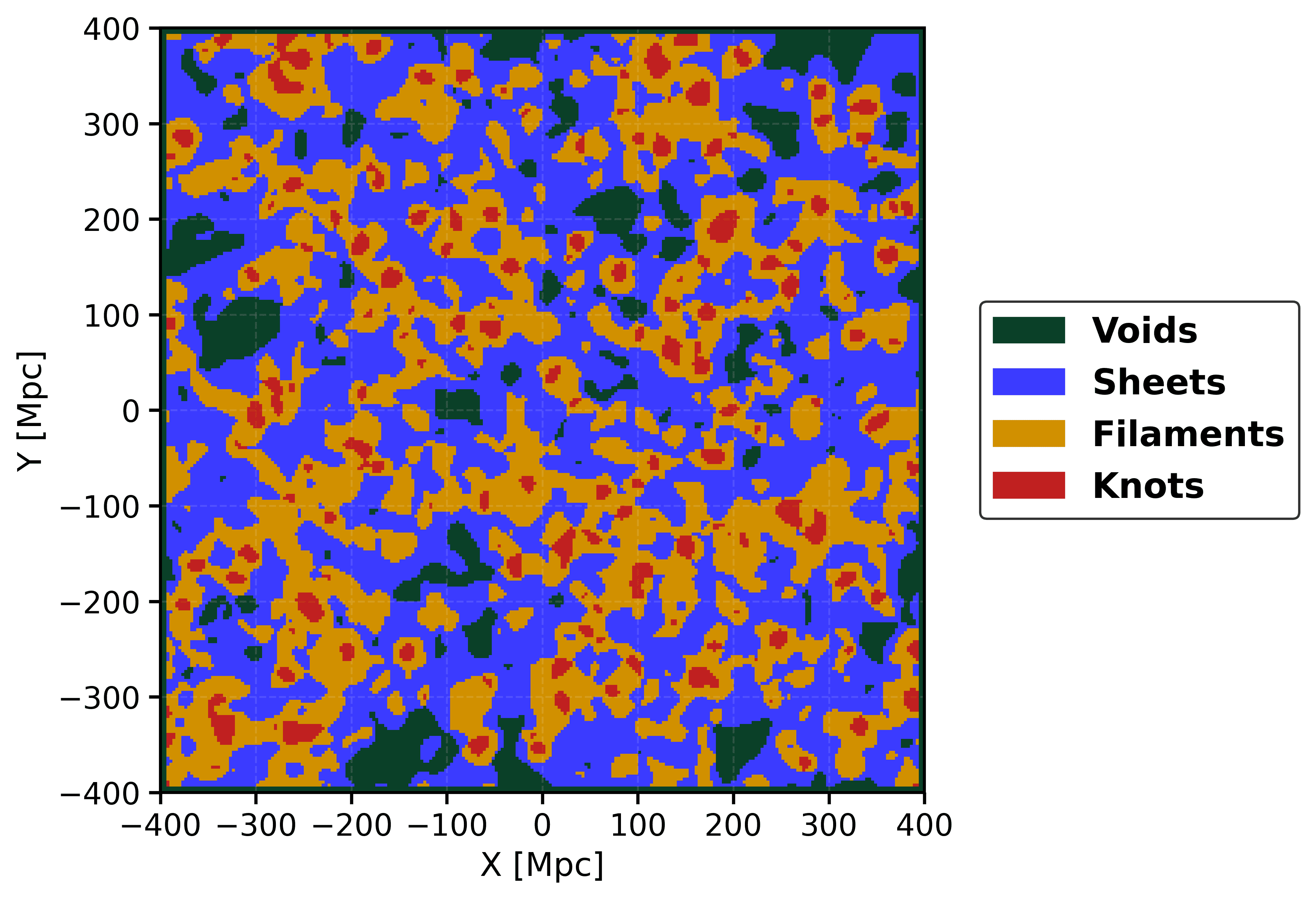}
\includegraphics[width=0.7\textwidth]{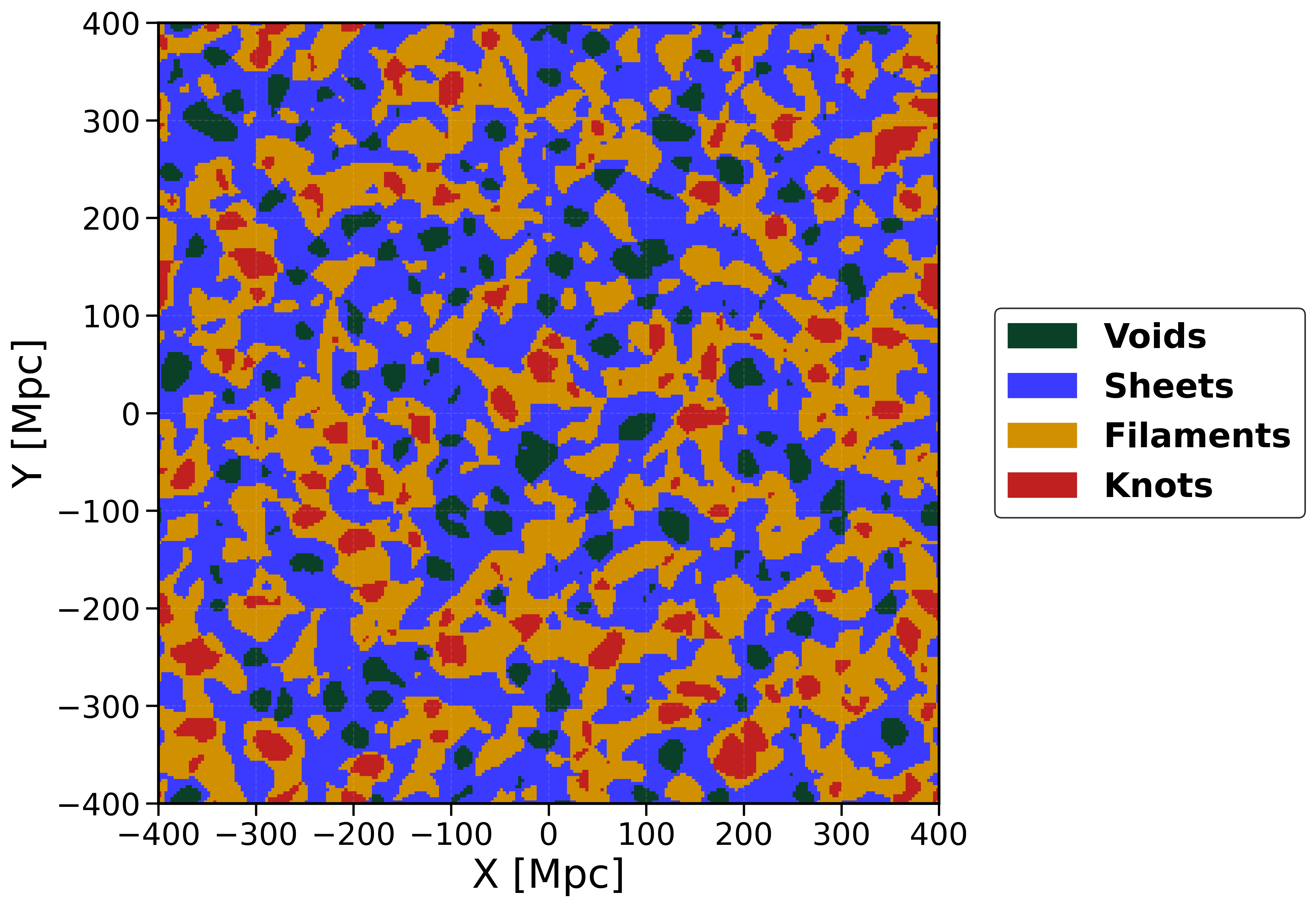}
\caption{Example slices of the T-Web environment maps for BGS (top, $R_s=10~{\rm Mpc}$), LRG (middle, $R_s=7~{\rm Mpc}$), and ELG (bottom, $R_s=10~{\rm Mpc}$). 
Colours indicate the four morphological environments: voids, sheets, filaments, and knots.}
\label{fig:tweb_slices}
\end{figure}

For each tracer we embed the galaxies in a cubic volume of side length $L_{\rm box}=800~{\rm Mpc}$ and discretize the overdensity field onto a $256^3$ Cartesian grid, corresponding to a cell size of
\[
\Delta x = \frac{L_{\rm box}}{256} = \frac{800~{\rm Mpc}}{256} \simeq 3.125~{\rm Mpc}.
\]
The T-Web classification is then computed after smoothing the density field with a Gaussian kernel of scale $R_s$. 
The ranges of smoothing scales explored for each tracer are summarized in Table~\ref{tab:tweb_representative}. For the illustrative slices shown in Figure~\ref{fig:tweb_slices}, we use the representative scales adopted in the fiducial comparison for each tracer: $R_s=10~{\rm Mpc}$ for BGS, $R_s=7~{\rm Mpc}$ for LRG, and $R_s=10~{\rm Mpc}$ for ELG. 
The fact that the displayed scales are not identical for all tracers reflects the tracer-dependent smoothing ranges explored in our analysis; in each case, the slice is shown at the representative scale used for the corresponding summary comparison below.

\FloatBarrier
\begin{table}[htbp]
\centering
\begin{tabular}{l|c|c|c|c|c}
\hline
Tracer & $R_s$ [Mpc] & Voids [\%] & Sheets [\%] & Filaments [\%] & Knots [\%] \\
\hline
BGS  & 10 & 15.27 & 45.64 & 38.45 & 5.49 \\
LRG  & 7 &  9.47 & 45.60 & 39.04 & 5.89 \\
ELG  & 10 &  7.40 & 47.10 & 39.20 & 6.30 \\
\hline
\end{tabular}
\caption{Volume fractions of cosmic-web environments for a representative smoothing scale of each tracer.
The complete set of values for all scales is provided in Appendix~\ref{app:tweb_tables}.}
\label{tab:tweb_representative}
\end{table}

Figure~\ref{fig:tweb_slices} shows example slices of the resulting environment maps for the three tracers, illustrating the visual appearance of the cosmic web as traced by different galaxy populations.

In Table~\ref{tab:tweb_representative} we list the volume fractions of the four environments obtained for a representative smoothing scale of each tracer (the full set of values for all scales is given in Appendix~\ref{app:tweb_tables}).
Sheets and filaments dominate the classified volume across all samples, while voids and knots contribute smaller fractions. 
The specific values vary with tracer and scale, reflecting differences in galaxy bias and redshift coverage.

\subsection{Identifying red and blue galaxies with Otsu's method}
\label{sec:otsu_method_general}

Galaxy colour distributions are typically bimodal, reflecting a blue star-forming population and a red quenched population.
To separate these components in a reproducible and data-driven manner, we adopt Otsu's thresholding method \citep{Otsu1979}, a variance-based classifier originally developed for image segmentation and recently used to characterize galaxy colour bimodality \citep{Pandey2022Otsu}.
For a given subsample (defined by tracer and redshift bin), the method selects the colour threshold that maximizes the separation between two classes while minimizing the dispersion within each class.

For each subsample we consider the colour $C \equiv (g-r)$ and construct a histogram with $M$ bins spanning a fixed colour range.
We use $(g-r)$ because it is uniformly available for all tracers in the DESI DR1 value-added products;
the divider is recomputed independently in each redshift bin to account for colour evolution with cosmic time. 
Let $n_i$ be the number of galaxies in bin $i$ and $
N=\sum_{i=1}^{M} n_i$ the total number of galaxies. The normalized probability in bin $i$ is
\begin{equation}
p_i=\frac{n_i}{N}, \qquad \sum_{i=1}^{M} p_i = 1 .
\end{equation}
Let $x_i$ denote the representative colour value of bin $i$ (e.g. the bin centre). 
A trial threshold at
bin $k$ partitions the sample into a ``blue'' class (bins $1,\dots,k$) and a ``red'' class (bins
$k+1,\dots,M$), with class probabilities
\begin{equation}
P_B(k)=\sum_{i=1}^{k} p_i \equiv w(k), \qquad
P_R(k)=\sum_{i=k+1}^{M} p_i = 1-w(k).
\end{equation}
The cumulative mean up to bin $k$ and the total mean are
\begin{equation}
\mu_k=\sum_{i=1}^{k} x_i p_i, \qquad
\mu_T=\sum_{i=1}^{M} x_i p_i ,
\end{equation}
and the class means are
\begin{equation}
\mu_B(k)=\frac{\mu_k}{w(k)}, \qquad
\mu_R(k)=\frac{\mu_T-\mu_k}{1-w(k)} .
\end{equation}
The within-class variances are
\begin{equation}
\sigma_B^2(k)=\frac{\sum_{i=1}^{k}(x_i-\mu_B)^2 p_i}{w(k)}, \qquad
\sigma_R^2(k)=\frac{\sum_{i=k+1}^{M}(x_i-\mu_R)^2 p_i}{1-w(k)} ,
\end{equation}
leading to the weighted within-class variance
\begin{equation}
\sigma_{wc}^2(k)=w(k)\sigma_B^2(k) + \left[1-w(k)\right]\sigma_R^2(k).
\end{equation}
Equivalently, one may maximize the between-class variance
\begin{equation}
\sigma_{bc}^2(k)=w(k)\left[1-w(k)\right]\left[\mu_B(k)-\mu_R(k)\right]^2 .
\end{equation}
Since the total variance satisfies $\sigma_T^2=\sigma_{wc}^2+\sigma_{bc}^2$ and is independent of
$k$, minimizing $\sigma_{wc}^2(k)$ is equivalent to maximizing $\sigma_{bc}^2(k)$. The optimal
threshold is therefore
\begin{equation}
k^\star=\arg\max_k \, \sigma_{bc}^2(k),
\end{equation}
which defines the colour divider $C_{\rm thr}$.

A single global colour cut is not appropriate when galaxies span a wide stellar-mass range, since both the red sequence and blue cloud shift systematically with $M_\star$. 
We therefore implement a \emph{mass-dependent} Otsu divider. 
In each redshift bin, galaxies are divided into stellar-mass intervals, and an Otsu threshold $C_{\rm thr}(M_\star)$ is computed independently in each interval (subject to a minimum galaxy count to ensure stability). 
The resulting discrete thresholds are interpolated to obtain a smooth divider in the $(g-r)$--$\log_{10}(M_\star)$ plane. Galaxies are classified as red if $(g-r) \ge C_{\rm thr}\!\left[\log_{10}(M_\star)\right]$,
and blue otherwise. 
This procedure is applied independently to each tracer (BGS, LRG, ELG) in each redshift bin.

\subsection{Colour--mass classification with a mass-dependent Otsu threshold}

We apply the mass-dependent Otsu procedure described in Section~\ref{sec:otsu_method_general} to the three DESI tracers. 
Figures~\ref{fig:bgs_otsu}--\ref{fig:elg_otsu} show the corresponding colour--mass planes in successive redshift bins. 
The black points mark the discrete Otsu thresholds measured in stellar-mass bins, while the black curves denote the interpolated mass-dependent red/blue dividers adopted for the final classification. 
This approach allows the separation to follow the observed evolution of the red sequence and blue cloud with stellar mass, rather than imposing a single global colour cut. 
The resulting number counts, split by tracer, redshift, and stellar mass, are listed in Appendix~\ref{app:otsu_counts}. After the quality cuts, the final samples contain 1,293,399 BGS galaxies ($0.15\le z<0.55$), 417,980 LRGs ($0.6\le z<0.9$), and 862,851 ELGs ($0.6\le z\le 1.6$).

\subsubsection{BGS colour--mass classification}

\begin{figure}[!htbp]
\centering
\includegraphics[width=0.7\textwidth]{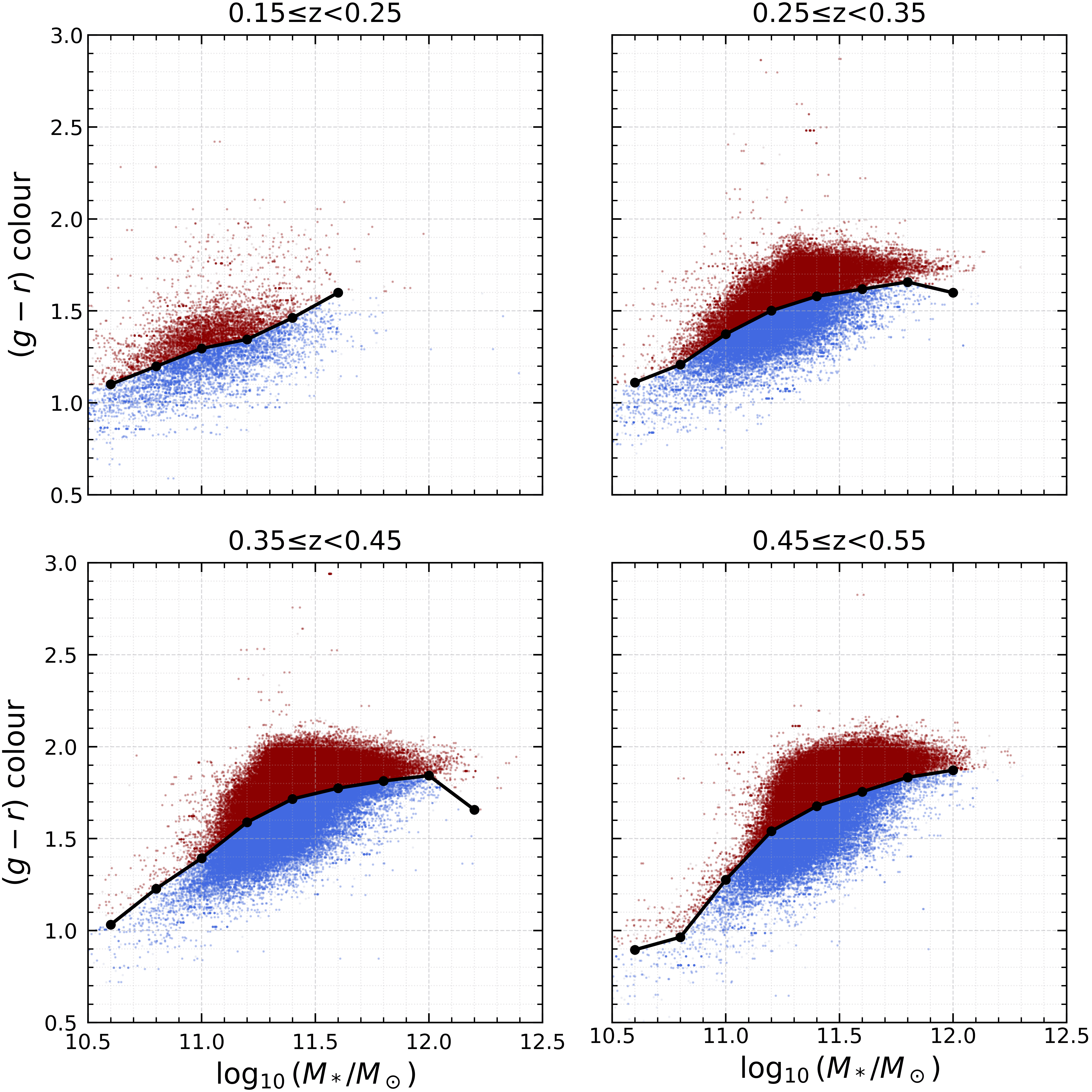}
\caption{Colour--mass diagrams for DESI DR1 BGS galaxies in four redshift bins spanning $0.15\leq z<0.55$. 
The black points show the Otsu thresholds measured in stellar-mass bins, and the black curve shows the interpolated mass-dependent divider used to classify galaxies into red and blue populations. 
The four redshift bins contain 37,849, 218,718, 667,696, and 366,617 galaxies, respectively.}
\label{fig:bgs_otsu}
\end{figure}

The BGS panels in Figure~\ref{fig:bgs_otsu} display the broadest colour--mass locus among the three tracers and show the clearest stellar-mass dependence of the divider. 
In all four redshift bins, the threshold shifts systematically toward redder $(g-r)$ colours with increasing stellar mass, while the low-mass population remains predominantly blue and the red sequence becomes progressively more prominent at higher masses. 
This behaviour is consistent with the increasing prevalence of quenched, red systems toward the massive end of the galaxy population. 
A second visible trend is the gradual depletion of the low-mass end toward the higher-redshift BGS bins, so that the highest-$z$ panels are increasingly dominated by massive systems. 
The appendix counts show the same tendency, with the high-mass subsample representing a large fraction of the total BGS sample.

\subsubsection{LRG colour--mass classification}

\begin{figure}[!htbp]
\centering
\includegraphics[width=1.0\textwidth]{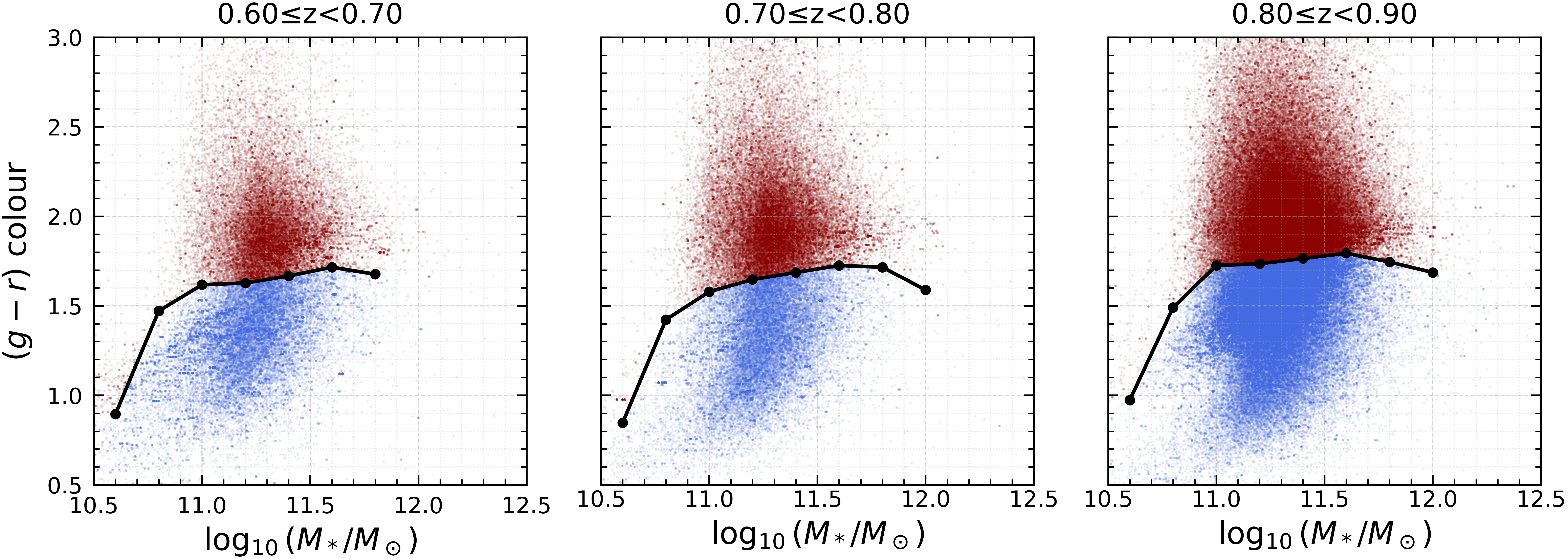}
\caption{Colour--mass diagrams for DESI DR1 LRGs in three redshift bins spanning $0.6\leq z<0.9$. 
The black points show the Otsu thresholds measured in stellar-mass bins, and the black curve shows the corresponding mass-dependent divider. 
The three redshift bins contain 62,144, 77,840, and 277,996 galaxies, respectively.}
\label{fig:lrg_otsu}
\end{figure}

\begin{figure}[!htbp]
\centering
\includegraphics[width=0.95\textwidth]{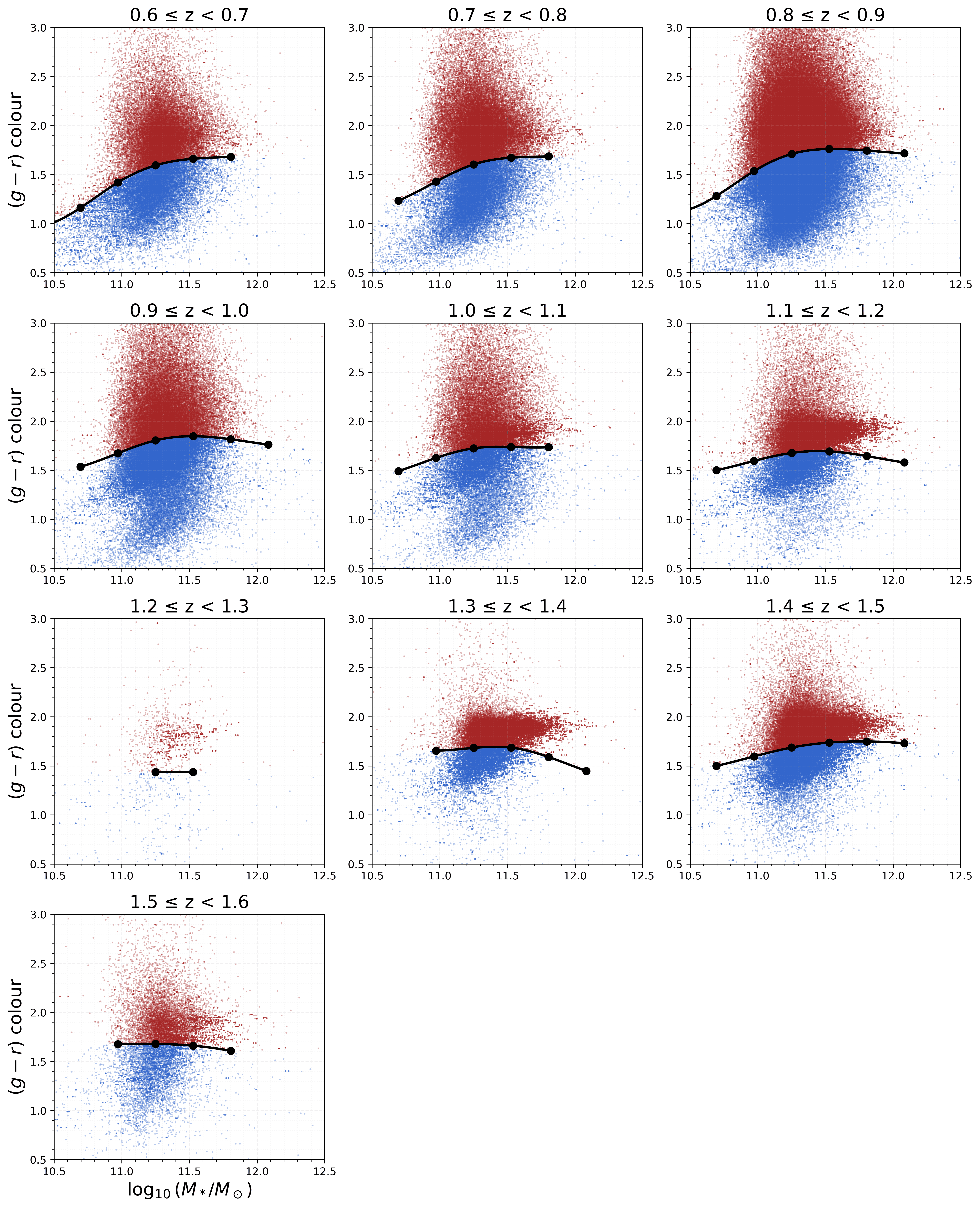}
\caption{Colour--mass diagrams for ELGs in 10 redshift bins from $z=0.6$ to 1.6. 
The black curve shows the mass-dependent Otsu divider derived in each redshift slice. 
The total sample contains 862,851 galaxies, with the richest bin at $0.8$--$0.9$ containing 273,025 objects and the sparsest at $1.2$--$1.3$ containing 1,887 objects.}
\label{fig:elg_otsu}
\end{figure}

The LRG sample, shown in Figure \ref{fig:lrg_otsu}, occupies a much narrower region of parameter space, concentrated almost entirely at the high-mass end of the distribution. 
Compared with BGS, the colour divider varies more weakly with stellar mass, especially above $\log_{10}(M_\star/M_\odot)\sim 11.2$, where the locus is dominated by a compact red population and only a relatively small blue tail remains. 
The main role of the mass-dependent threshold in this case is therefore not to track a broad transition across the full star-forming population, but to capture the residual colour structure within a tracer that is already strongly weighted toward massive red galaxies. 
The LRG panels also show that the red sequence remains well defined across the full redshift range considered here, with only modest evolution in the location of the divider.

\subsubsection{ELG colour--mass classification}

The ELG sample from Figure \ref{fig:elg_otsu} extends the same classification scheme to a tracer whose spectroscopic completeness is closely tied to emission-line detectability. 
Consequently, the inferred divider should be interpreted as separating redder and bluer ELGs within the observed sample, rather than as a completeness-independent boundary identical in character to that obtained for BGS or LRGs. 
The sparse population in the $1.2\le z<1.3$ bin should be treated with particular caution. 
A likely contributing factor is redshift-dependent spectroscopic incompleteness, since secure ELG redshifts rely heavily on the detection of the $[\mathrm{O\,II}]\,\lambda 3726,3729$ doublet and the success rate is known to vary when the observed doublet overlaps strong night-sky emission features. 
For this reason, the reduced occupancy of that bin is more conservatively described as likely affected by observational selection, rather than interpreted as a purely astrophysical deficit of star-forming galaxies.

\section{Results}\label{sec:results}

In this section we investigate how the red (quenched) and blue (star-forming) galaxy populations depend on cosmic-web environment for the DESI DR1 samples analyzed in this work (BGS, LRG, and ELG).
Throughout, we characterise environment using the T-Web classification into four morphological types: voids, sheets, filaments, and knots, adopting the smoothing scales listed in subsection \ref{sec:tweb_method_general} for each tracer.

We present red and blue fractions as functions of redshift and stellar mass, and we additionally quantify how the global red and blue populations are partitioned across the cosmic web using relative fractions.

For a given environment, we define the red fraction (RF) and blue fraction (BF) as
\begin{equation}
\mathrm{RF} \equiv \frac{n_R}{n_R+n_B},\qquad
\mathrm{BF} \equiv \frac{n_B}{n_R+n_B},
\label{eq:rf_bf}
\end{equation}
where $n_R$ and $n_B$ denote the numbers of red and blue galaxies in that environment, respectively.
By construction, $\mathrm{RF}+\mathrm{BF}=1$, so these quantities provide a direct measure of the relative importance of quenched versus star-forming systems in each web morphology at fixed redshift and/or stellar mass.
The red/blue split is determined using the mass-dependent Otsu threshold in the colour–mass plane, applied independently in each redshift bin for each tracer, following the methodology of subsection~\ref{sec:otsu_method_general}.
\newpage
To quantify how red and blue galaxies are distributed across the full cosmic web, we also compute the relative red fraction (RRF) and relative blue fraction (RBF) in each environment $i$ as
\begin{equation}
\mathrm{RRF}_i \equiv \frac{(n_R)_i}{\sum_{j=1}^{4}(n_R)_j},\qquad
\mathrm{RBF}_i \equiv \frac{(n_B)_i}{\sum_{j=1}^{4}(n_B)_j},
\label{eq:rrf_rbf}
\end{equation}
where $(n_R)_i$ and $(n_B)_i$ are the red and blue galaxy counts in the $i$th environment, and the sum over $j$ runs over the four web types (void, sheet, filament, knot).
Unlike RF and BF, which compare red and blue populations \emph{within} the same environment, RRF and RBF measure the contribution of each environment to the \emph{total} red or blue population at a given redshift and/or stellar mass. 
In particular, $\sum_i \mathrm{RRF}_i = 1$ and $\sum_i \mathrm{RBF}_i = 1$ by definition, but $\mathrm{RRF}_i$ does not determine $\mathrm{RBF}_i$ because red and blue populations can be distributed differently across the cosmic web.

Using these metrics, we examine: 
(i) the redshift evolution of RF/BF in each environment,
(ii) the stellar-mass dependence of RF/BF within each environment, and 
(iii) the evolution and mass dependence of RRF/RBF, which together highlight how the locations of quenched and star-forming galaxies shift across voids, sheets, filaments, and knots over cosmic time for each DESI tracer.

\subsection{Redshift evolution of red and blue fractions}
\label{sec:rf_bf_z}

\begin{figure}[!htbp]
\centering
\includegraphics[width=0.9\textwidth]{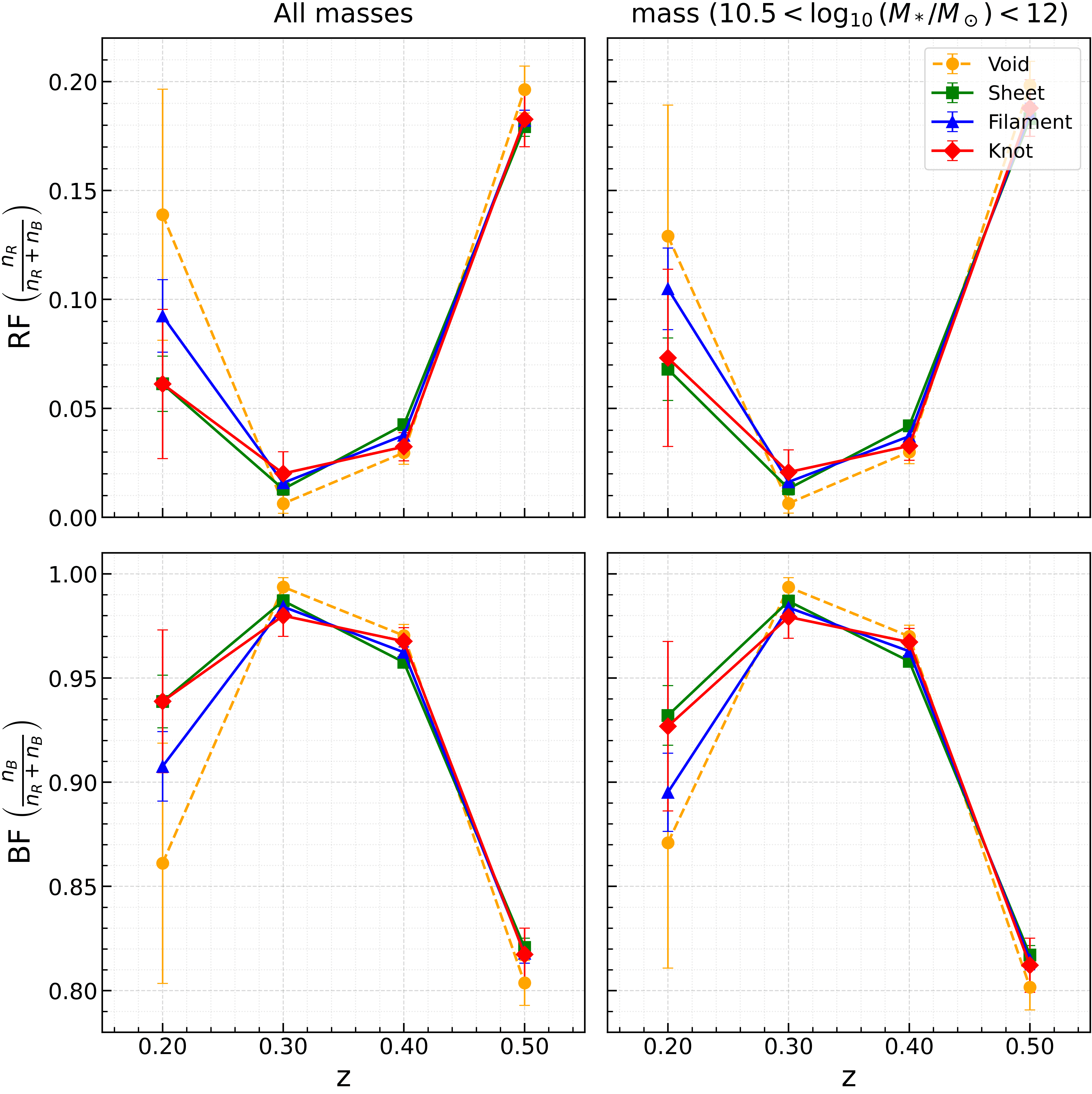}
\caption{Red fraction (RF, top panels) and blue fraction (BF, bottom panels) as functions of redshift for the BGS sample in different cosmic-web environments. The left panels correspond to the full sample, and the right panels to the high-mass subsample with $10.5 < \log_{10}(M_\ast/M_\odot) < 12$. Error bars denote the $1\sigma$ binomial uncertainties. For clarity, the vertical axis ranges are truncated to $0.00$--$0.22$ in the RF panels and to $0.80$--$1.00$ in the BF panels.}
\label{fig:bgs_rf_bf_z}
\end{figure}

\begin{figure}[!htbp]
\centering
\includegraphics[width=0.9\textwidth]{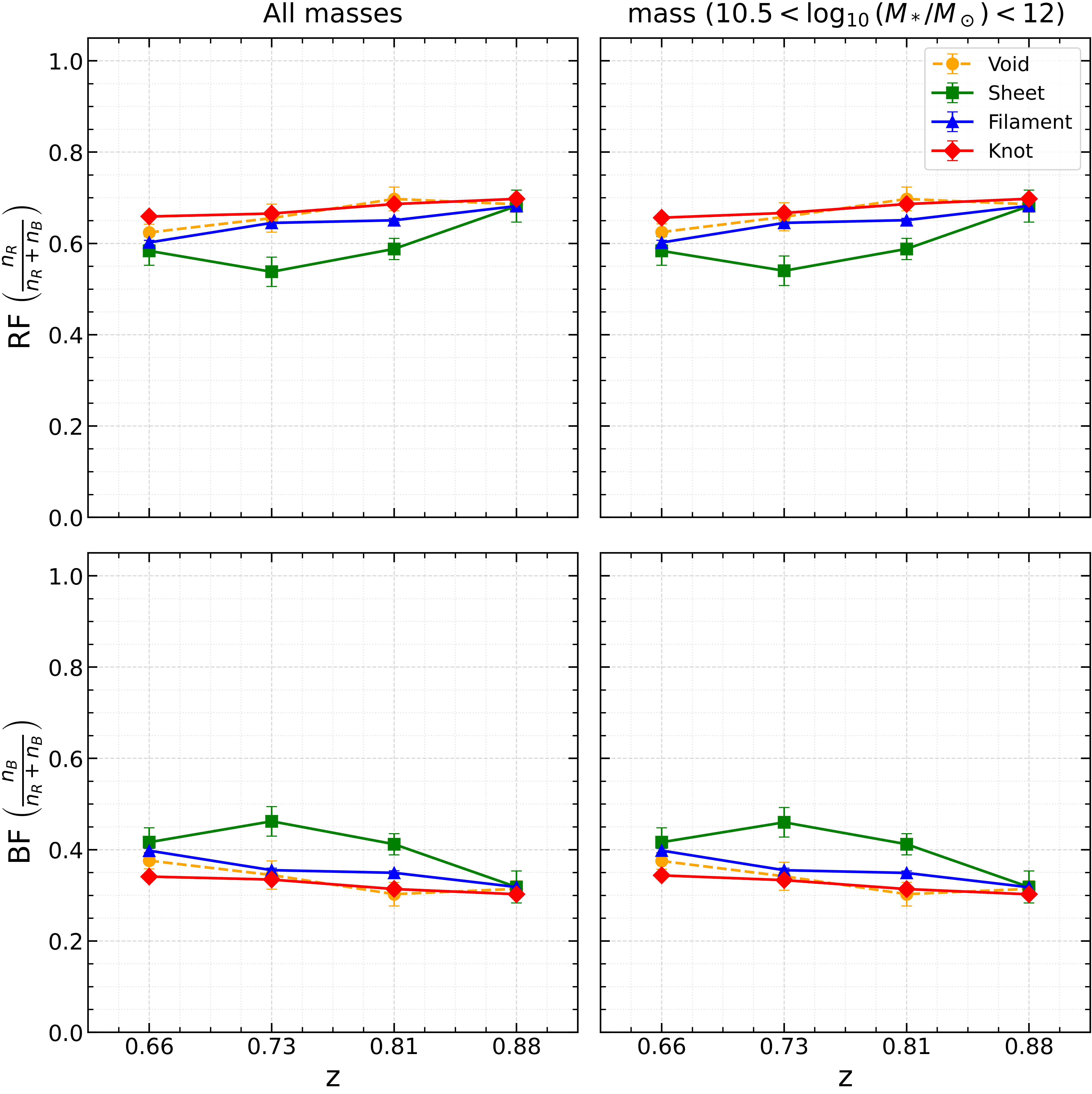}
\caption{Red fraction (RF, top panels) and blue fraction (BF, bottom panels) as functions of redshift in different cosmic-web environments for the LRG sample. The left column shows the full sample, while the right column shows the high-mass subsample with $10.5 < \log_{10}(M_*/M_\odot) < 12$. Error bars denote the $1\sigma$ binomial uncertainties.}
\label{fig:lrg_rf_bf_z}
\end{figure}

\begin{figure}[!htbp]
\centering
\includegraphics[width=0.9\textwidth]{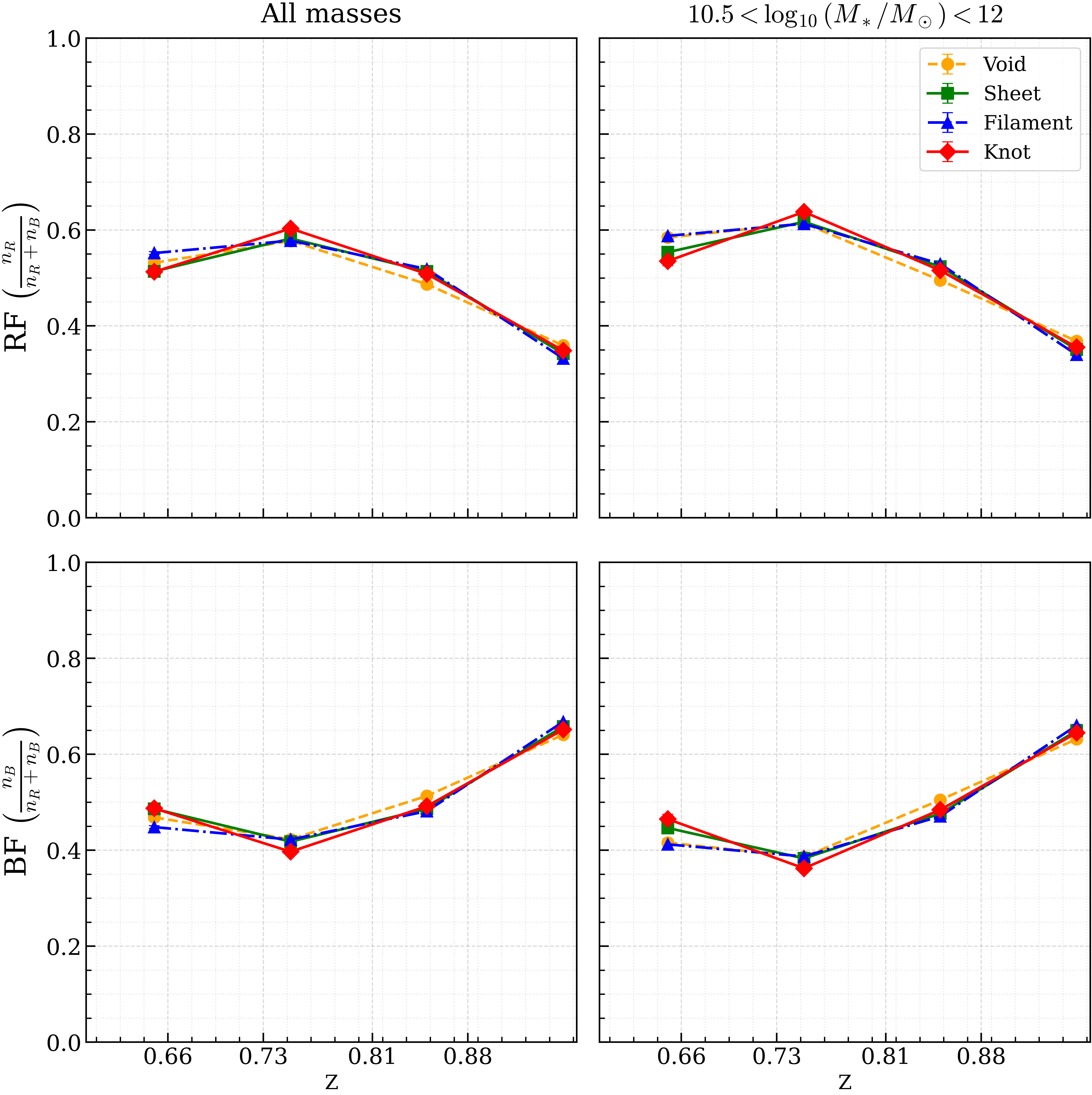}
\caption{Red fraction as a function of redshift for ELG galaxies (high-mass subsample, $10.5<\log_{10}(M_\ast/M_\odot)<12$) at $0.6\leq z<0.95$ in four cosmic-web environments. The left column shows the full sample, while the right column shows the high-mass subsample with $10.5 < \log_{10}(M_*/M_\odot) < 12$. Error bars represent $1\sigma$ binomial uncertainties.}
\label{fig:elg_rf_bf_z}
\end{figure}

We first examine how the red and blue fractions evolve with redshift in different cosmic-web environments. Figure~\ref{fig:bgs_rf_bf_z} shows the results for the BGS sample over $0.15\leq z<0.55$, Figure~\ref{fig:lrg_rf_bf_z} for LRGs over $0.6\leq z<0.9$, and Figure~\ref{fig:elg_rf_bf_z} for ELGs over $0.65\leq z\leq0.95$.

The BGS sample (Fig.~\ref{fig:bgs_rf_bf_z}) shows a clear non-monotonic evolution of the red fraction (RF) with redshift in all environments. At $z\sim0.20$, the RF is $13.89\pm5.76\%$ in voids, $6.13\pm1.27\%$ in sheets, $9.24\pm1.66\%$ in filaments, and $6.12\pm3.42\%$ in knots. It then drops sharply at $z\sim0.30$ to $0.63\pm0.44\%$, $1.29\pm0.29\%$, $1.58\pm0.33\%$, and $2.01\pm0.99\%$ in voids, sheets, filaments, and knots, respectively. At $z\sim0.40$, the RF rises slightly to $2.97\pm0.53\%$ in voids, $4.25\pm0.27\%$ in sheets, $3.77\pm0.26\%$ in filaments, and $3.24\pm0.65\%$ in knots. Toward the highest-redshift bin, $z\sim0.50$, the RF increases markedly in all environments, reaching $19.63\pm1.08\%$ in voids, $17.93\pm0.44\%$ in sheets, $18.22\pm0.47\%$ in filaments, and $18.27\pm1.26\%$ in knots. The complementary blue fraction (BF) follows the opposite trend, decreasing from $86.11\pm5.76\%$, $93.87\pm1.27\%$, $90.76\pm1.66\%$, and $93.88\pm3.42\%$ at $z\sim0.20$ to $80.37\pm1.08\%$, $82.07\pm0.44\%$, $81.78\pm0.47\%$, and $81.73\pm1.26\%$ at $z\sim0.50$ in voids, sheets, filaments, and knots, respectively. Overall, the environmental differences remain small compared with the redshift evolution, and the ordering is not stable across bins.

The LRG sample (Fig.~\ref{fig:lrg_rf_bf_z}) is characterized by systematically high red fractions in all environments and a moderate increase with redshift. In the full sample, at $z\approx0.66$ the RF is $65.90\pm0.45\%$ in knots, $62.40\pm1.81\%$ in voids, $60.21\pm0.48\%$ in filaments, and $58.37\pm3.15\%$ in sheets. At $z\approx0.73$, the RF becomes $66.54\pm0.45\%$ in knots, $65.55\pm3.08\%$ in voids, $64.49\pm0.47\%$ in filaments, and $53.78\pm3.23\%$ in sheets. At $z\approx0.81$, the highest RF is found in voids, $69.75\pm2.59\%$, followed by knots, $68.62\pm0.39\%$, filaments, $65.07\pm0.42\%$, and sheets, $58.80\pm2.32\%$. In the highest-redshift bin, $z\approx0.88$, the RF is $69.76\pm0.49\%$ in knots, $68.58\pm1.76\%$ in voids, $68.18\pm0.52\%$ in filaments, and $68.18\pm3.51\%$ in sheets. Thus, while knots are generally among the reddest environments and sheets tend to show lower RF at intermediate redshift, the environmental ordering is not strictly fixed across all bins; in particular, at $z\approx0.88$ the values for voids, filaments, and sheets are very similar within the quoted uncertainties. The corresponding blue fractions follow the complementary trend, decreasing from $41.63\pm3.15\%$--$34.10\pm0.45\%$ at $z\approx0.66$ to about $31\%$ at $z\approx0.88$ in all environments.

The ELG full sample exhibits a pronounced redshift evolution in the red fraction (RF), while the environmental dependence remains comparatively weak. In the full sample (Fig.~\ref{fig:elg_rf_bf_z}, left column), the RF at \(z \approx 0.65\) ranges from \(51.27 \pm 0.95\%\) in knots and \(51.38 \pm 0.27\%\) in sheets to \(53.16 \pm 0.76\%\) in voids and \(55.18 \pm 0.31\%\) in filaments. It then rises at \(z \approx 0.75\), reaching \(57.73 \pm 0.72\%\), \(58.21 \pm 0.24\%\), \(57.78 \pm 0.28\%\), and \(60.30 \pm 0.63\%\) in voids, sheets, filaments, and knots, respectively. At \(z \approx 0.85\), the RF decreases to values close to \(50\%\) in all environments, namely \(48.69 \pm 0.40\%\) in voids, \(51.36 \pm 0.14\%\) in sheets, \(51.86 \pm 0.15\%\) in filaments, and \(50.83 \pm 0.31\%\) in knots. By \(z \approx 0.95\), the RF drops further to \(35.92 \pm 0.66\%\), \(34.22 \pm 0.23\%\), \(33.22 \pm 0.21\%\), and \(34.85 \pm 0.61\%\), respectively. Thus, in the full ELG sample, the dominant behaviour is a strong redshift dependence, whereas the environmental ordering is mild and not preserved across bins.

In the high-mass subsample (Fig.~\ref{fig:elg_rf_bf_z}, right column), the same qualitative trend is present, but the RF is systematically higher than in the full sample. At \(z \approx 0.65\), the RF is \(58.43 \pm 0.80\%\) in voids, \(55.36 \pm 0.28\%\) in sheets, \(58.78 \pm 0.32\%\) in filaments, and \(53.51 \pm 0.98\%\) in knots. At \(z \approx 0.75\), it increases further to \(61.40 \pm 0.74\%\), \(61.70 \pm 0.25\%\), \(61.29 \pm 0.28\%\), and \(63.77 \pm 0.64\%\), respectively. At \(z \approx 0.85\), the RF again declines to \(49.52 \pm 0.40\%\) in voids, \(52.36 \pm 0.14\%\) in sheets, \(52.92 \pm 0.15\%\) in filaments, and \(51.56 \pm 0.31\%\) in knots, before falling to \(36.83 \pm 0.67\%\), \(35.08 \pm 0.23\%\), \(33.99 \pm 0.21\%\), and \(35.54 \pm 0.62\%\) at \(z \approx 0.95\). Therefore, selecting the high-mass ELGs shifts the RF upward in every redshift bin, but does not alter the overall non-monotonic redshift evolution or the relatively weak environmental separation.

The blue fraction (BF) follows the expected complementary behaviour. In the full sample, it decreases from \(46.84 \pm 0.76\%\), \(48.62 \pm 0.27\%\), \(44.82 \pm 0.31\%\), and \(48.73 \pm 0.95\%\) at \(z \approx 0.65\) to \(42.27 \pm 0.72\%\), \(41.79 \pm 0.24\%\), \(42.22 \pm 0.28\%\), and \(39.70 \pm 0.63\%\) at \(z \approx 0.75\), before increasing to \(51.31 \pm 0.40\%\), \(48.64 \pm 0.14\%\), \(48.14 \pm 0.15\%\), and \(49.17 \pm 0.31\%\) at \(z \approx 0.85\), and finally to \(64.08 \pm 0.66\%\), \(65.78 \pm 0.23\%\), \(66.78 \pm 0.21\%\), and \(65.15 \pm 0.61\%\) at \(z \approx 0.95\), in voids, sheets, filaments, and knots, respectively. The high-mass subsample shows the same complementary trend, with correspondingly lower BF values in each bin. Overall, the ELG results indicate that redshift evolution is the dominant effect, while stellar mass raises the RF systematically and the environmental modulation remains secondary.

\subsection{Stellar-mass dependence of the red fraction}
\label{sec:rf_bf_mass}

\begin{figure}[!htbp]
\centering
\includegraphics[width=0.9\textwidth]{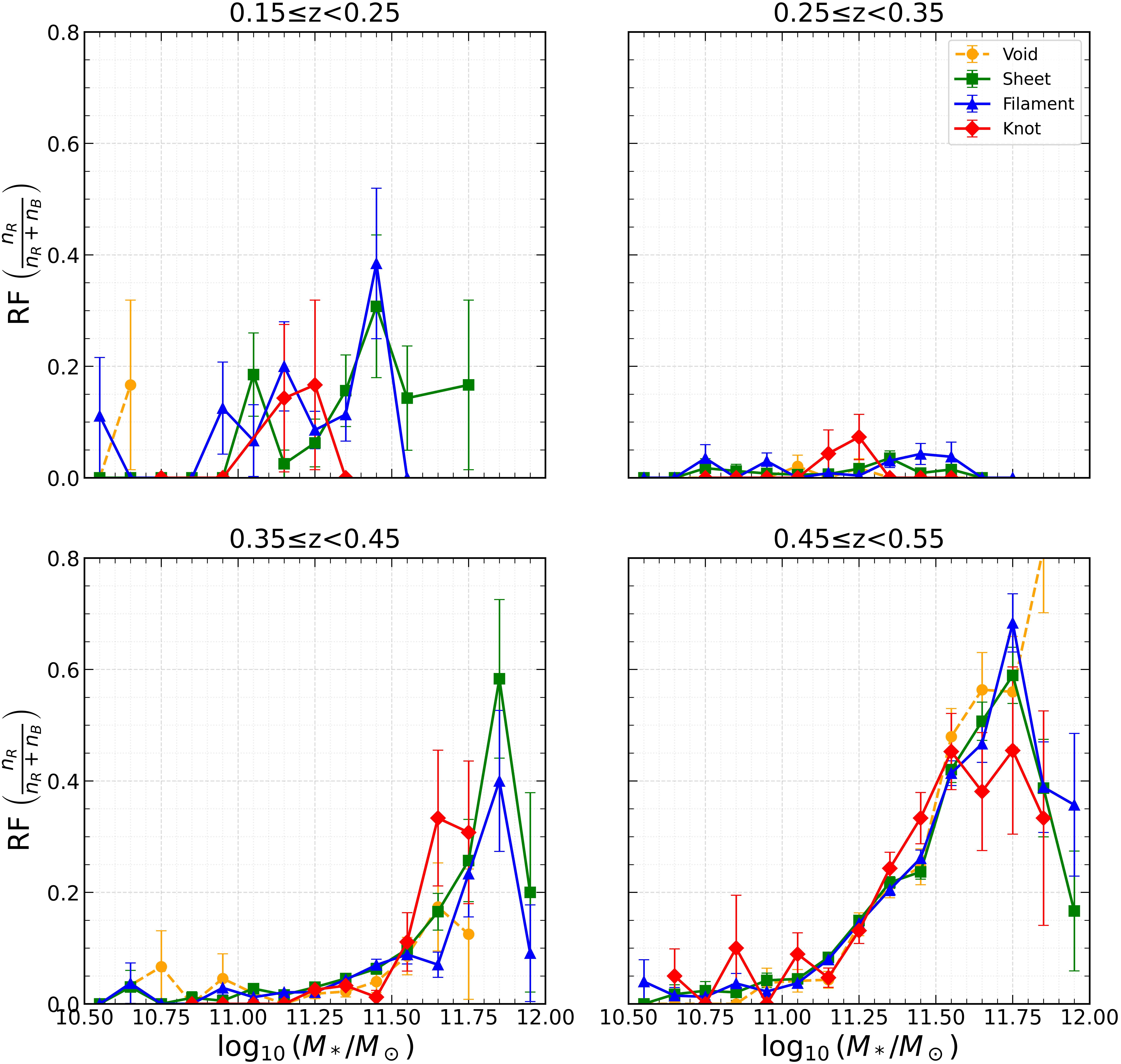}
\caption{Red fraction as a function of stellar mass for BGS galaxies in four cosmic-web environments, shown in four redshift bins: $0.15\leq z<0.25$, $0.25\leq z<0.35$, $0.35\leq z<0.45$, and $0.45\leq z<0.55$. Error bars represent $1\sigma$ binomial uncertainties.}
\label{fig:bgs_rf_mass}
\end{figure}

\begin{figure}[!htbp]
\centering
\includegraphics[width=0.8\textwidth]{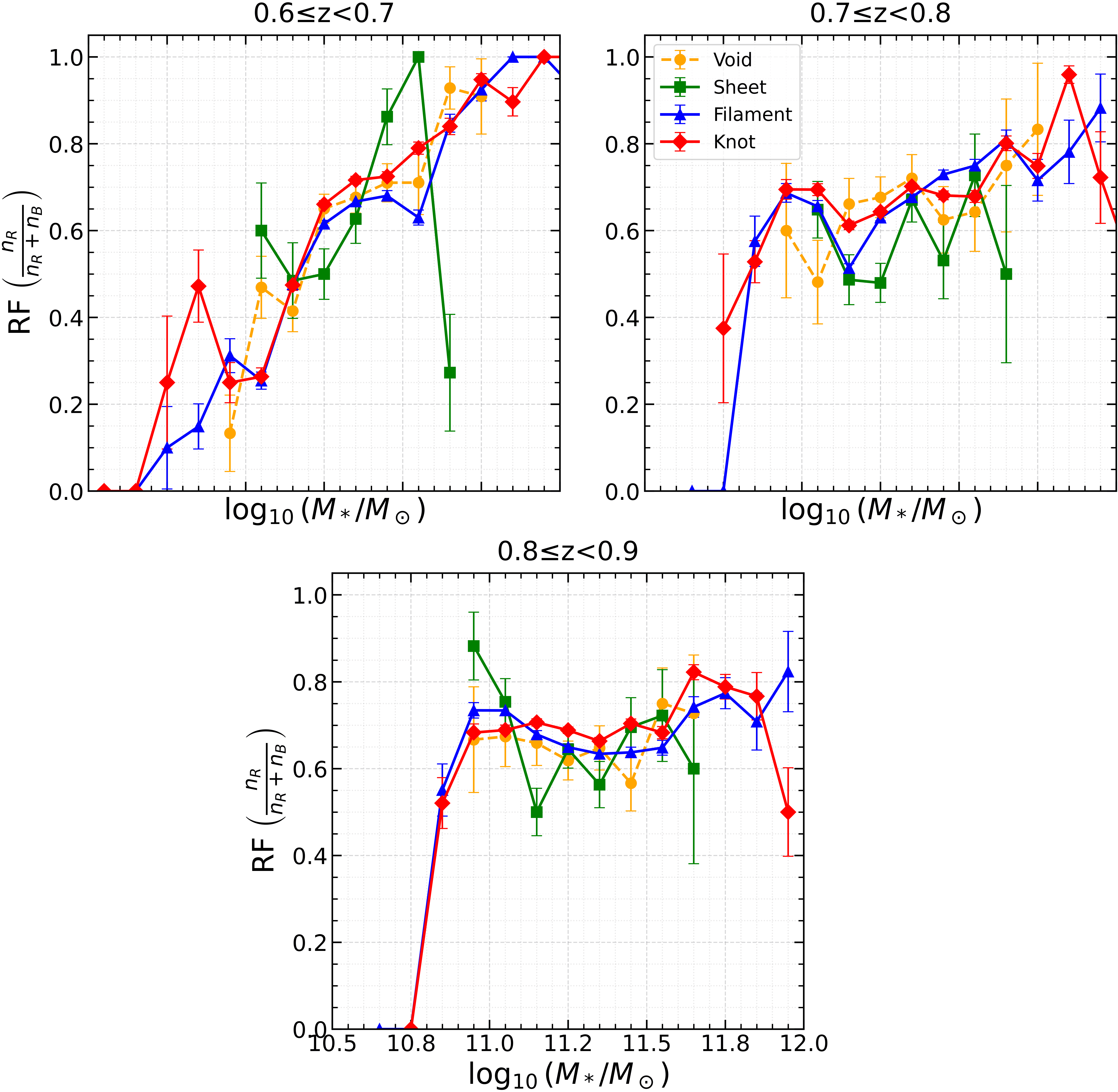}
\caption{Red fraction as a function of stellar mass for LRG galaxies at $0.6\leq z<0.9$ in four cosmic-web environments. Error bars represent $1\sigma$ binomial uncertainties.}
\label{fig:lrg_rf_mass}
\end{figure}

\begin{figure}[!htbp]
\centering
\includegraphics[width=0.8\textwidth]{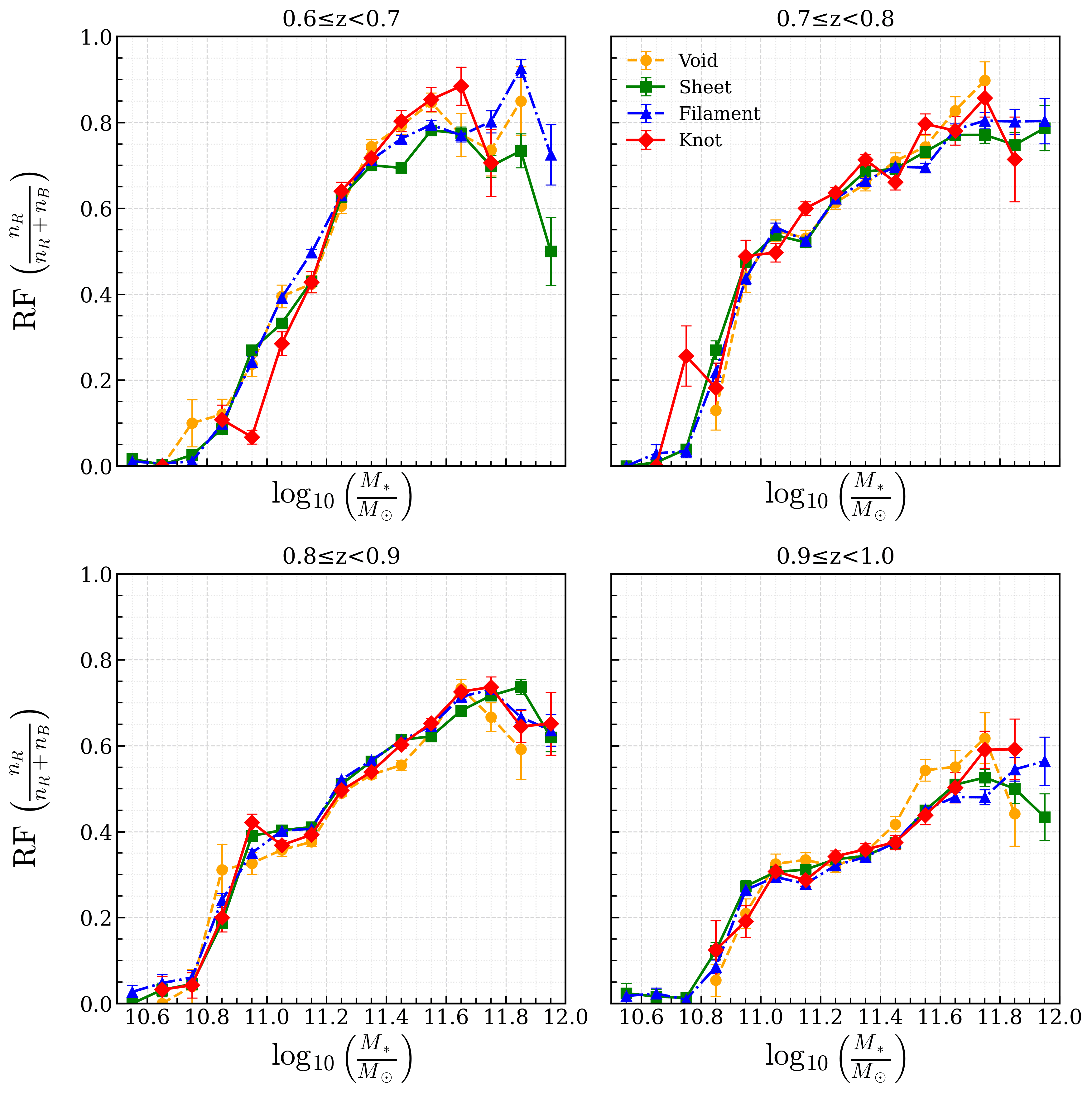}
\caption{Red fraction as a function of stellar mass, $\log_{10}(M_\star/M_\odot)$, for ELG galaxies in the Void, Sheet, Filament, and Knot environments. The four panels correspond to the redshift ranges $0.60 \leq z < 0.70$, $0.70 \leq z < 0.80$, $0.80 \leq z < 0.90$, and $0.90 \leq z < 1.00$. Error bars represent $1\sigma$ binomial uncertainties.}
\label{fig:elg_rf_mass}
\end{figure}

We next examine how the red fraction varies with stellar mass for each tracer. The corresponding results are shown in Figure~\ref{fig:bgs_rf_mass} for BGS, Figure~\ref{fig:lrg_rf_mass} for LRGs, and Figure~\ref{fig:elg_rf_mass} for ELGs. A common trend is evident across all tracers and environments: at fixed redshift, the red fraction increases with stellar mass. This behaviour is consistent with the increasing importance of quenching in massive galaxies, where internal feedback and halo-related processes are expected to be more effective. Since the blue fraction is simply the complement of the red fraction, $\mathrm{BF}=1-\mathrm{RF}$, we do not show a separate BF--mass figure; its behaviour follows directly from the RF trends discussed below.

The BGS panels (Fig.~\ref{fig:bgs_rf_mass}) show the clearest rise of RF toward the high-mass end. In all environments, the relation steepens noticeably at $\log_{10}(M_\ast/M_\odot)\gtrsim 11.2$, which may be regarded as an approximate transition mass above which red galaxies become increasingly dominant. At fixed stellar mass, filaments and knots generally reach larger RF values than voids and sheets, especially in the higher-redshift bins where the upturn becomes more pronounced. The stronger point-to-point fluctuations seen in the lowest-redshift bin, particularly in voids and knots, are most likely driven by the smaller effective galaxy counts in those bins rather than by a change in the overall trend. The corresponding blue fraction therefore decreases with stellar mass, mirroring the increase in RF.

The LRG sample (Fig.~\ref{fig:lrg_rf_mass}) occupies a regime that is already dominated by massive, predominantly red galaxies, so the mass dependence is smoother and the RF remains high over most of the plotted range. The environmental separation nevertheless becomes more apparent toward lower redshift. In the $0.8\leq z<0.9$ bin, knots lie slightly above filaments and sheets across much of the mass range, while voids generally show the lowest RF. By $0.6\leq z<0.7$, the environmental contrast is strongest: knots consistently attain the highest RF, exceeding $\sim80\%$ at $\log_{10}(M_\ast/M_\odot)\gtrsim11.5$, whereas voids remain at the low end, with RF values of about $\sim50\%$ near $\log_{10}(M_\ast/M_\odot)\approx10.8$. This pattern suggests that environmental effects act mainly by enhancing red-galaxy dominance within a tracer that is already strongly biased toward massive systems. As expected, the blue fraction correspondingly declines toward higher stellar mass.

The ELG panels (Fig.~\ref{fig:elg_rf_mass}), show that, in the high-mass sample, the red fraction (RF) increases systematically with stellar mass in every redshift interval. This trend is clearest in the well-populated mass range, roughly $10.9 \lesssim \log_{10}(M_\ast/M_\odot) \lesssim 11.7$, where the statistical uncertainties are small and the RF rises smoothly toward higher masses in all environments. Thus, even within the ELG population, more massive galaxies are progressively more likely to belong to the red population.

The environmental dependence is present but comparatively modest, and it does not follow a single fixed ordering across all redshift and mass bins. At low redshift, the curves for sheets, filaments, and knots often lie close together at intermediate and high masses, while voids can be slightly lower in some bins but overlap with the denser environments in others. At higher redshift, the same overall mass trend remains, but the separation among environments becomes weaker and more variable. In particular, the retained bins indicate substantial overlap between the different cosmic-web environments once the low-count edge bins are excluded.

Overall, the ELG result indicates that stellar mass is the primary driver of the transition toward red populations, while the cosmic-web environment plays a secondary role that modulates the RF without imposing a universally monotonic ranking among voids, sheets, filaments, and knots. The blue fraction follows as the complementary quantity, decreasing with stellar mass in all redshift bins.

\subsection{Relative fractions: partitioning of red and blue populations across the cosmic web}
\label{sec:rrf_rbf}

We now examine how the total red and blue populations are partitioned across the cosmic web using the relative red fraction (RRF) and relative blue fraction (RBF). These quantities measure the contribution of each environment to the total red or blue population, independently of the relative abundance of red and blue galaxies within each environment. In contrast to the volume fractions of environments (which describe how much space each morphology occupies), RRF and RBF answer a different question: where do the red and blue galaxies actually live?

\subsubsection{Redshift evolution of relative fractions}

\begin{figure}[!htbp]
\centering
\includegraphics[width=0.8\textwidth]{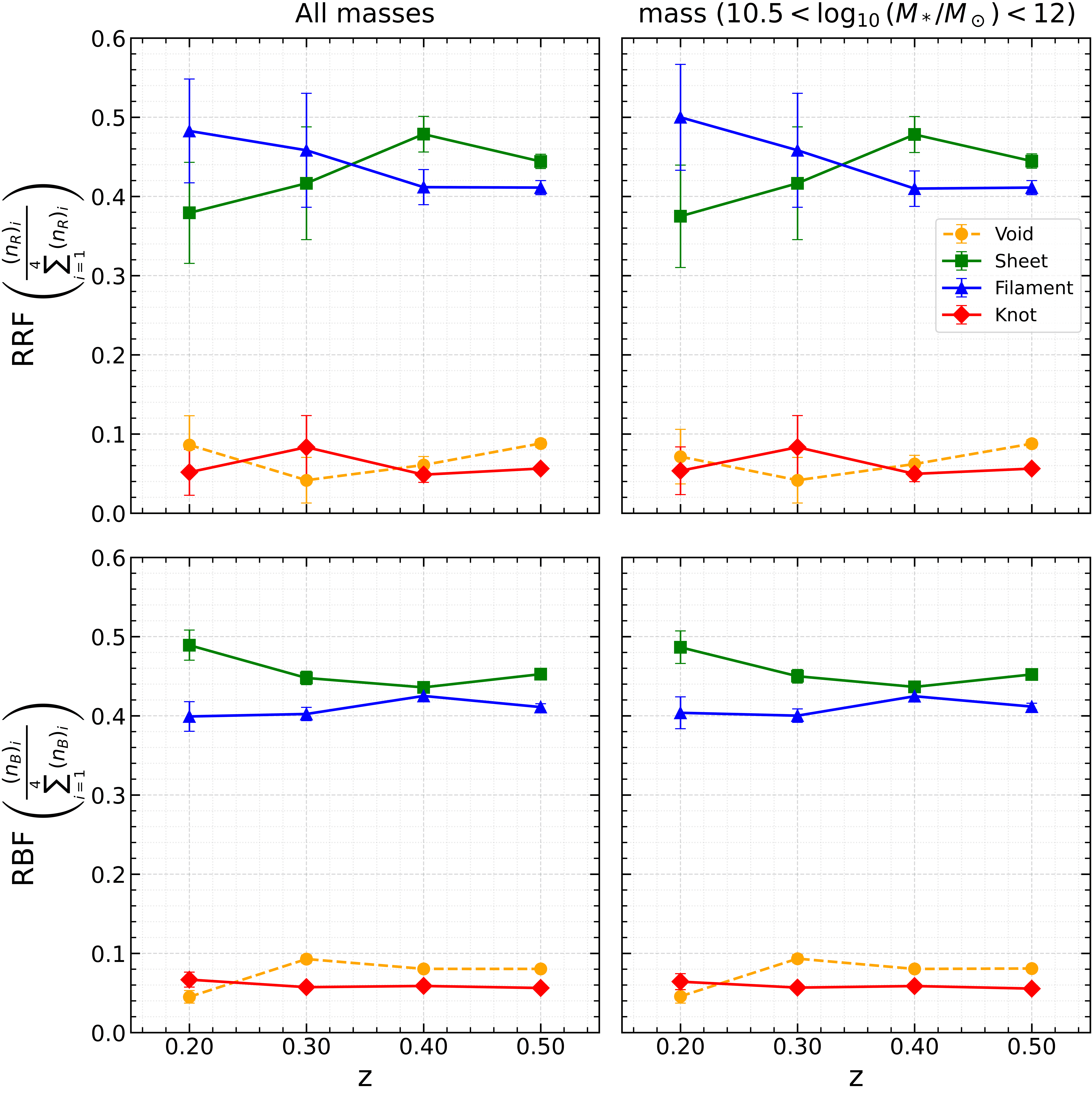}
\caption{(Top) Relative red fraction (RRF) as a function of redshift for BGS galaxies in four cosmic-web environments. (Bottom) Relative blue fraction (RBF) as a function of redshift for BGS galaxies. Error bars denote $1\sigma$ binomial uncertainties.}
\label{fig:bgs_rrf_rbf_z}
\end{figure}

\begin{figure}[!htbp]
\centering
\includegraphics[width=0.8\textwidth]{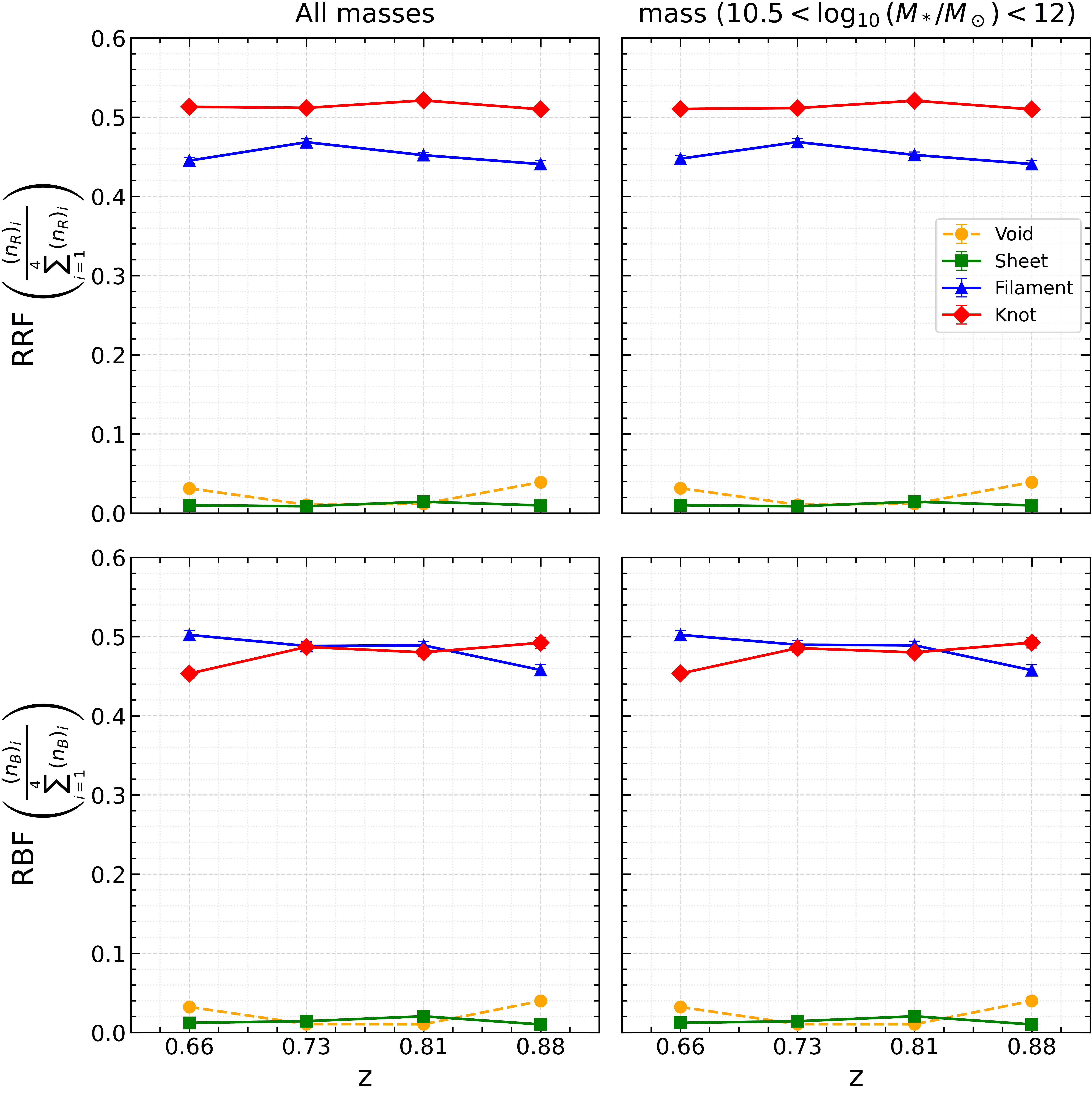}
\caption{(Top) Relative red fraction (RRF) as a function of redshift for LRG galaxies. (Bottom) Relative blue fraction (RBF) as a function of redshift for LRG galaxies. Error bars denote $1\sigma$ binomial uncertainties.}
\label{fig:lrg_rrf_rbf_z}
\end{figure}

\begin{figure}[!htbp]
\centering
\includegraphics[width=0.8\textwidth]{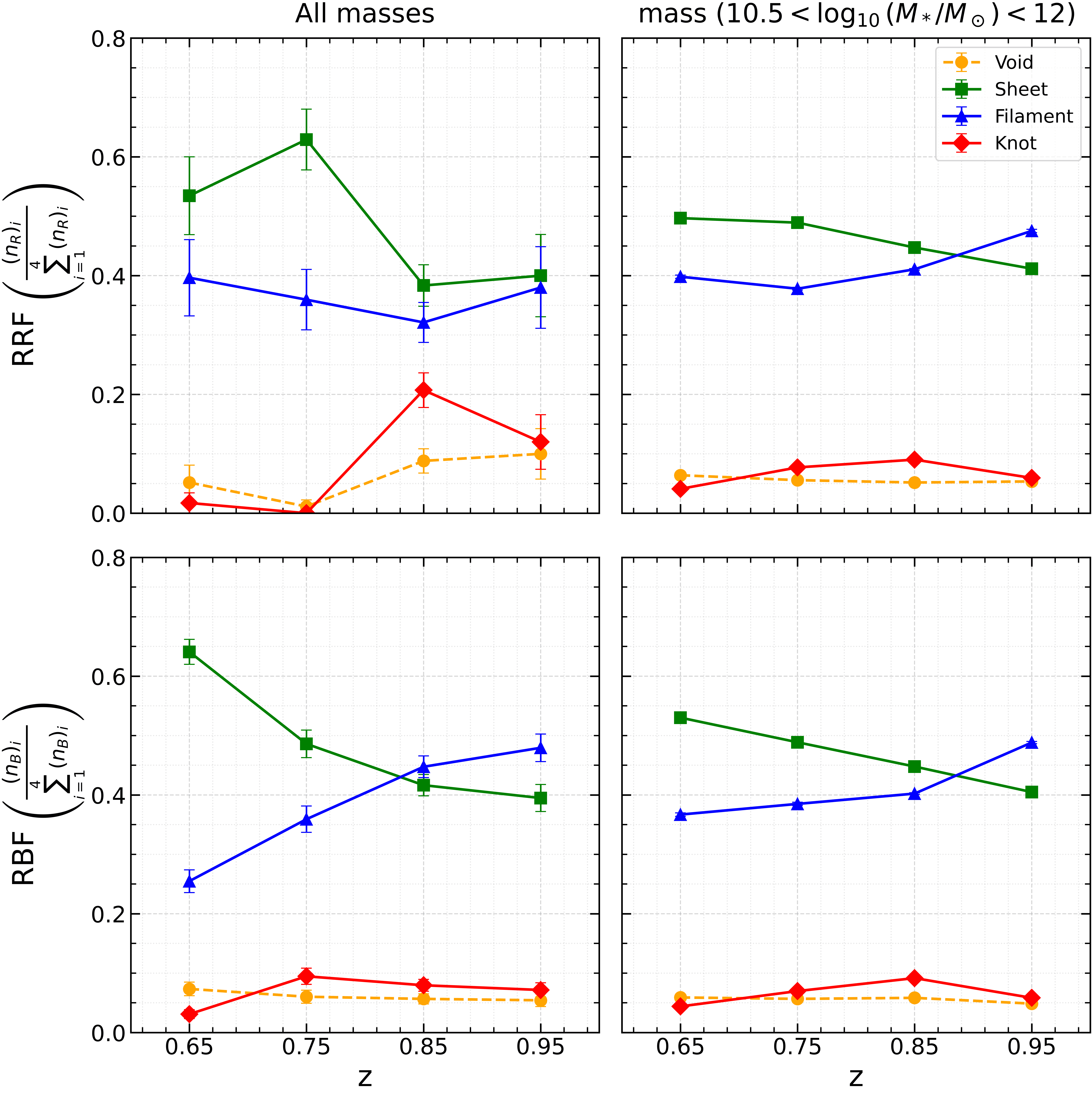}
\caption{(Top) Relative red fraction (RRF) as a function of redshift for ELG galaxies (full sample). (Bottom) Relative blue fraction (RBF) as a function of redshift for ELG galaxies. Error bars denote $1\sigma$ binomial uncertainties.}
\label{fig:elg_rrf_rbf_z}
\end{figure}

Relative fractions provide a complementary view of galaxy evolution by tracking how the total red and blue populations are partitioned among the different cosmic-web environments at each epoch. In contrast to the ordinary red and blue fractions, which measure the internal composition of a given environment, RRF and RBF emphasize how the global red and blue populations are distributed across the web. This makes them particularly useful for identifying where quenched and star-forming systems are preferentially hosted as a function of redshift. Figures~\ref{fig:bgs_rrf_rbf_z}–\ref{fig:elg_rrf_rbf_z} present the corresponding trends for the three DESI tracers.

In the BGS sample, the dominant share of both red and blue galaxies is associated with sheets and filaments throughout the full redshift interval. Together, these two environments account for approximately $\sim 60$--$80\%$ of the red population and $\sim 70$--$90\%$ of the blue population over $0.15\le z<0.55$, indicating that intermediate-density structures are the main reservoirs in which galaxies are distributed at low redshift. 
The relative red fraction in sheets increases from $z\sim0.2$ to $z\sim0.4$ and then declines slightly, while the filament contribution shows a modest decrease over the same range. 
By contrast, knots and voids remain subdominant contributors. The knot contribution to the red population rises from about $\sim 5\%$ at $z\sim0.1$ to nearly $\sim 10\%$ at $z\sim0.2$, whereas voids typically contribute less than $\sim 10\%$ to both populations.
The knot fraction is also systematically larger for red galaxies than for blue galaxies, suggesting that the densest environments host a somewhat enhanced share of more evolved systems. 
Overall, the blue population follows a similar environmental partition, but with a weaker knot contribution, consistent with ongoing star formation being less concentrated in the highest-density regions.

The LRG results show a markedly different partition. Here the relative fractions are dominated by filaments and knots, whereas sheets and voids make only a minor contribution to both the red and blue populations. Together, filaments and knots host roughly $\sim 95\%$ of the red population and a similarly dominant fraction of the blue population across the full redshift range, while sheets and voids each remain at only a few percent. The red population is especially concentrated in knots, which contribute about $\sim 51$--$52\%$, while filaments provide the second-largest share at roughly $\sim 44$--$47\%$. Blue LRGs, though much less numerous in an absolute sense, are again distributed primarily between filaments and knots, each contributing approximately $\sim 46$--$50\%$ depending on redshift. This behaviour is consistent with the fact that LRGs are intrinsically biased toward massive systems inhabiting dense large-scale environments. Thus, even the residual blue component of the LRG sample is largely confined to the denser parts of the cosmic web. The similarity between the full-sample and high-mass panels further indicates that this environmental partition is not driven by low-mass objects, but reflects the underlying structure of the tracer itself.

For ELG Figure~\ref{fig:elg_rrf_rbf_z}, the relative fractions show that sheets and filaments host the dominant share of both red and blue galaxies in both the all-mass and high-mass samples. In the all-mass case, the red relative fraction (RRF) is initially dominated by sheets, which contribute about \(0.53\) at \(z\approx0.65\) and \(0.63\) at \(z\approx0.75\), while filaments contribute about \(0.40\) and \(0.36\), respectively. At higher redshift, however, the sheet contribution declines and the filament contribution becomes comparatively more important, with the two becoming nearly comparable by \(z\approx0.95\). A similar trend is seen in the high-mass sample, where the sheet RRF decreases from about \(0.50\) at \(z\approx0.65\) to \(0.41\) at \(z\approx0.95\), while the filament RRF rises from about \(0.40\) to \(0.48\), overtaking sheets in the highest-redshift bin.

The blue relative fraction (RBF) follows the same broad picture. Sheets dominate at low redshift, contributing about \(0.64\) in the all-mass sample and \(0.53\) in the high-mass sample at \(z\approx0.65\), whereas the filament contribution increases steadily with redshift and exceeds that of sheets by \(z\approx0.95\). Voids remain a minor component throughout, typically contributing only \(\sim5\)–\(10\%\). Knots are usually subdominant as well, but they are not negligible in all cases: their red relative fraction becomes noticeably enhanced at intermediate redshift, especially in the all-mass sample around \(z\approx0.85\). Overall, the ELG population is preferentially distributed along the sheet--filament network, with a redshift-dependent shift from sheet-dominated to increasingly filament-dominated relative fractions.

\FloatBarrier

\subsubsection{Stellar mass dependence of relative fractions}

\begin{figure}[!htbp]
\centering
\includegraphics[width=0.75\textwidth]{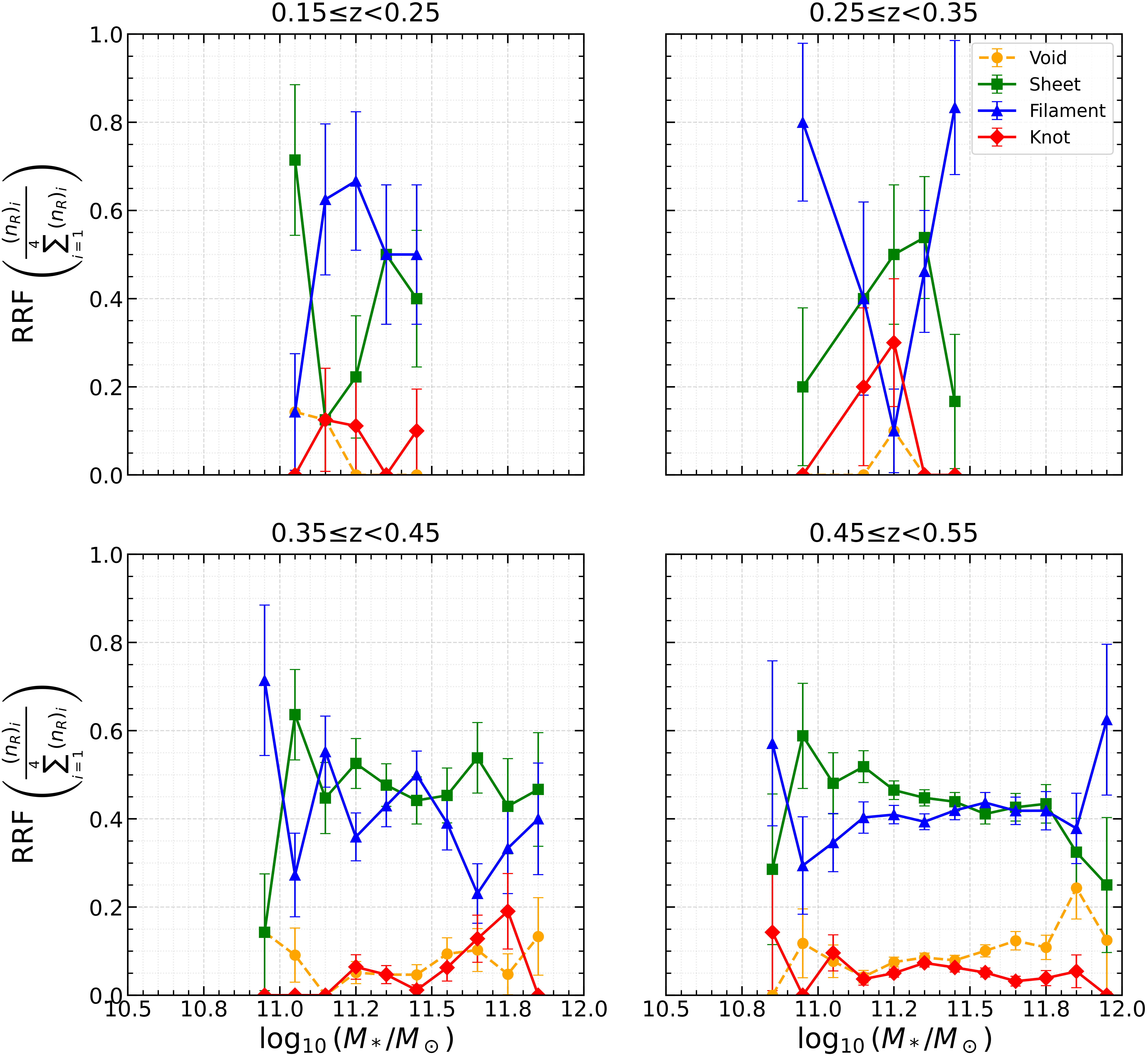}
\includegraphics[width=0.75\textwidth]{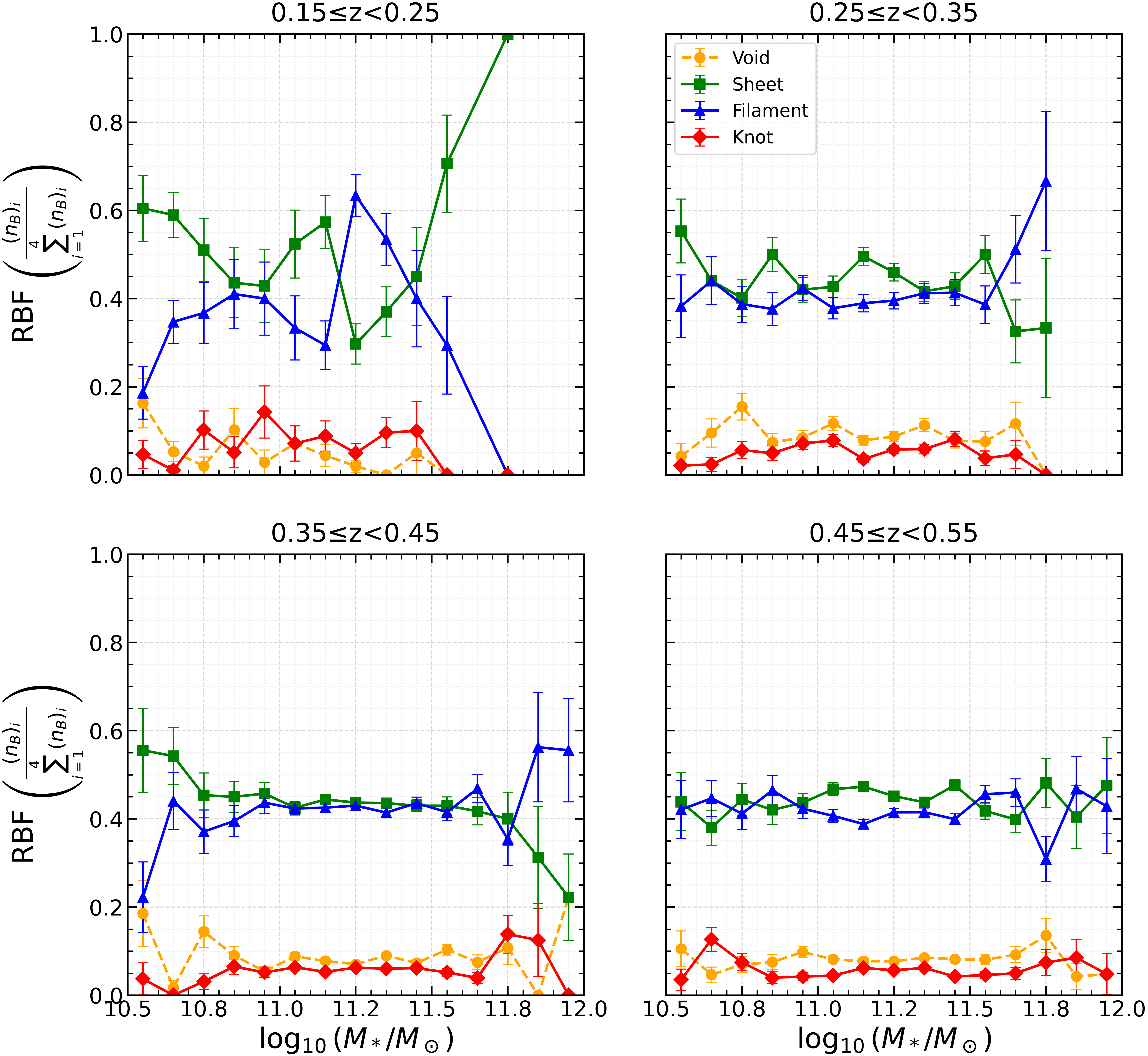}
\caption{(Top) Relative red fraction (RRF) as a function of stellar mass for BGS galaxies at $0.15\leq z<0.55$ in four cosmic-web environments. (Bottom) Relative blue fraction (RBF) as a function of stellar mass for BGS galaxies. Error bars represent $1\sigma$ binomial uncertainties in both panels.}
\label{fig:bgs_rrf_rbf_mass}
\end{figure}

\begin{figure}[!htbp]
\centering
\includegraphics[width=0.73\textwidth]{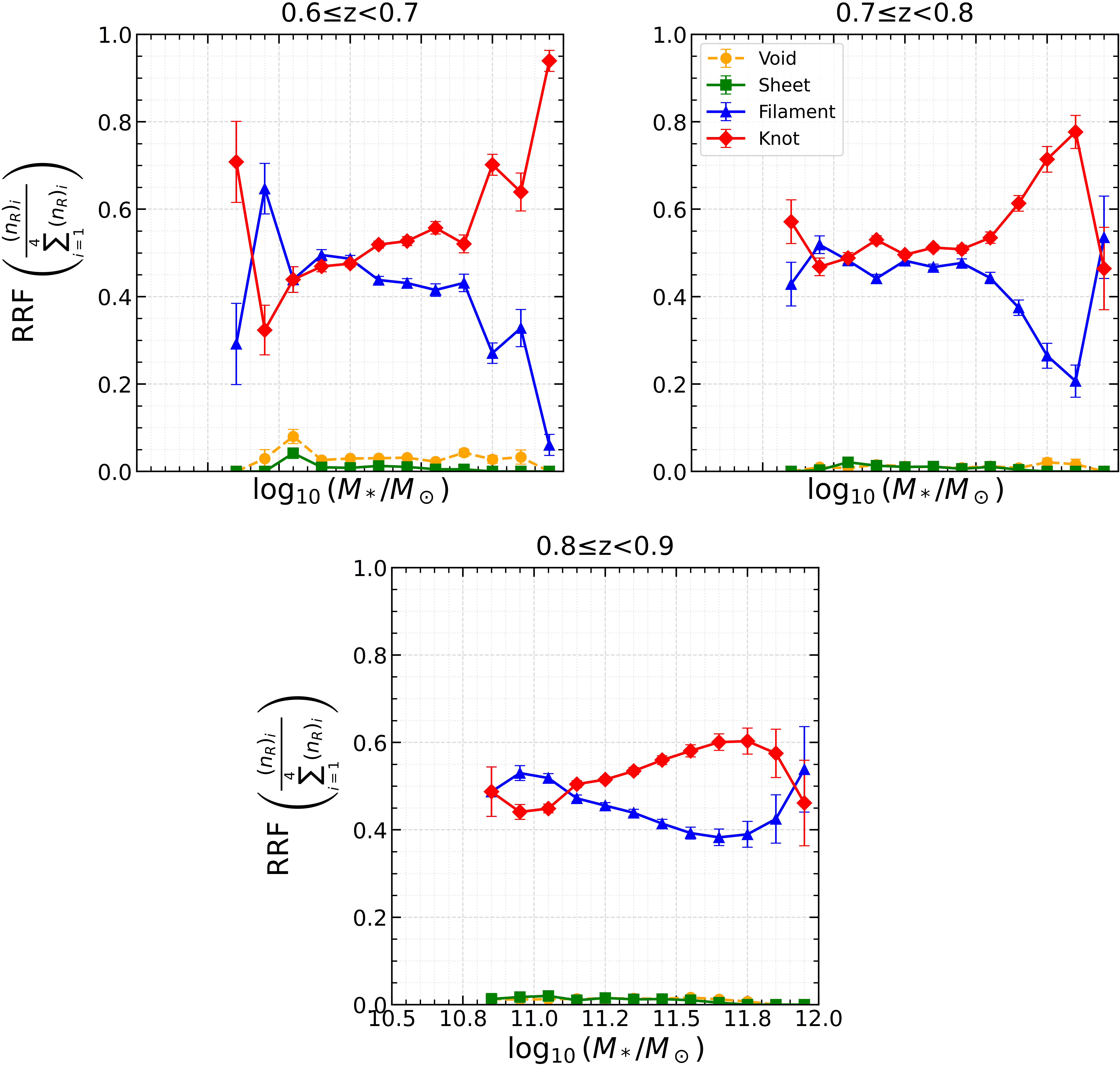}
\includegraphics[width=0.73\textwidth]{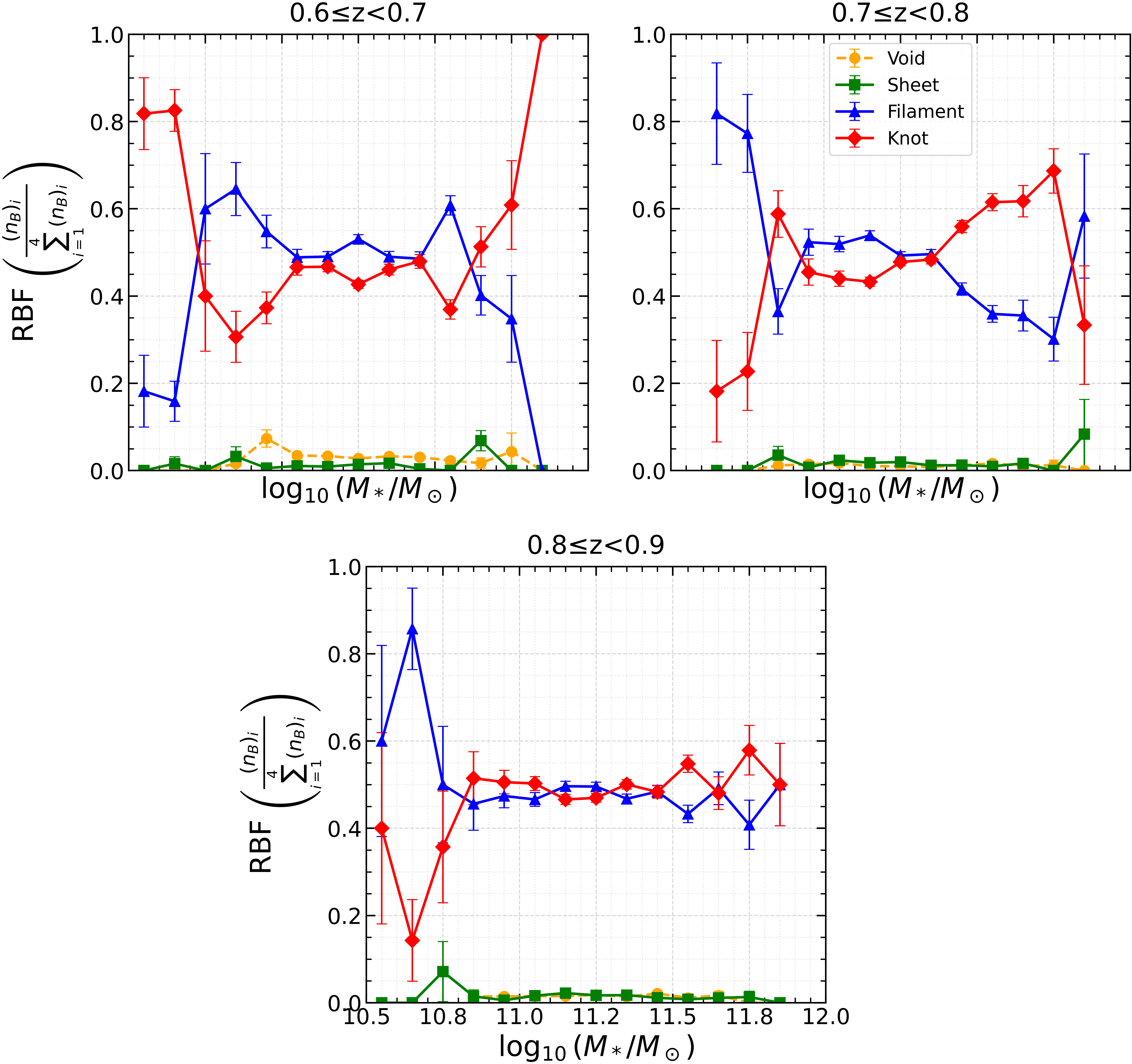}
\caption{(Top) Relative red fraction (RRF) as a function of stellar mass for LRG galaxies at $0.6\leq z<0.9$ in four cosmic-web environments. (Bottom) Relative blue fraction (RBF) as a function of stellar mass for LRG galaxies. Error bars represent $1\sigma$ binomial uncertainties in both panels.}
\label{fig:lrg_rrf_rbf_mass}
\end{figure}

\begin{figure}[!htbp]
\centering
\includegraphics[width=0.7\textwidth]{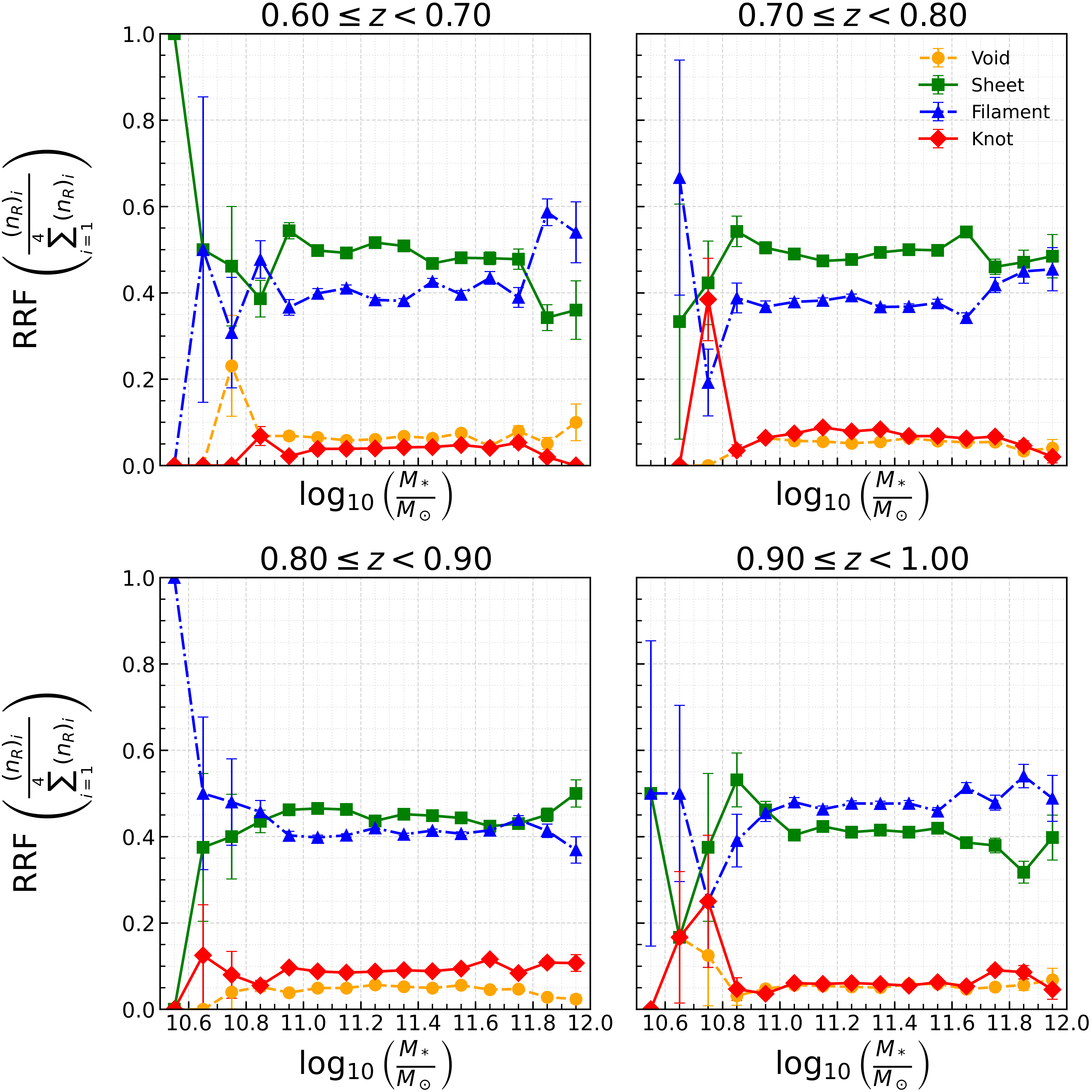}
\includegraphics[width=0.7\textwidth]{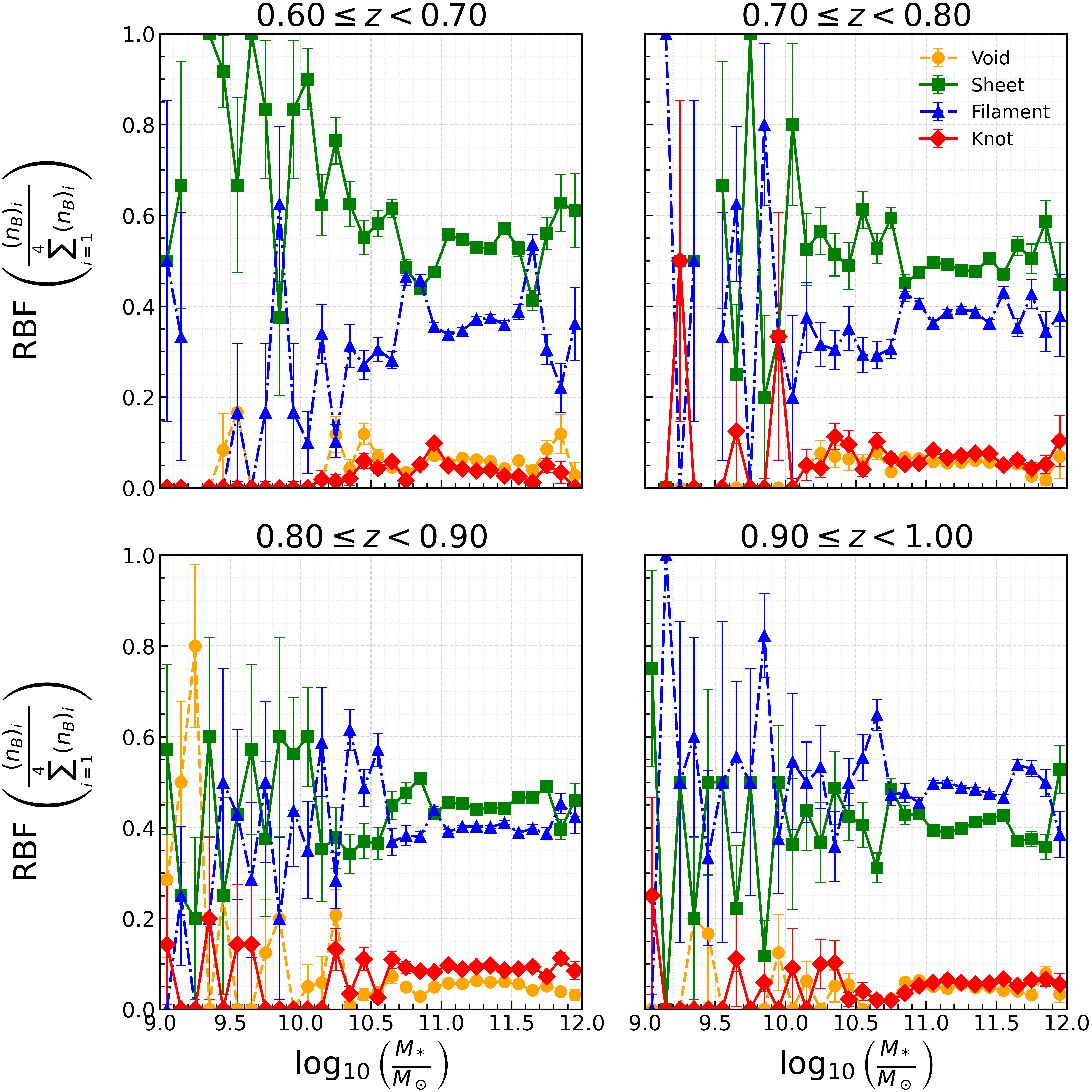}
\caption{(Top) Relative red fraction (RRF) as a function of stellar mass for ELG galaxies (high-mass subsample) at $0.6\leq z<0.9$ in four cosmic-web environments. (Bottom) Relative blue fraction (RBF) as a function of stellar mass for ELG galaxies. Error bars represent $1\sigma$ binomial uncertainties in both panels.}
\label{fig:elg_rrf_rbf_mass}
\end{figure}

Relative fractions provide a complementary description of the galaxy population by showing how the total red and blue samples at a given stellar mass are partitioned among the different cosmic-web environments. In contrast to the ordinary red and blue fractions, which characterize the internal composition of a single environment, RRF and RBF trace where the red and blue populations are preferentially located across the web. Figures~\ref{fig:bgs_rrf_rbf_mass}--\ref{fig:elg_rrf_rbf_mass} show this mass dependence for the three DESI tracers.

A broad common pattern is apparent. In BGS and ELG, sheets and filaments generally host the largest shares of both red and blue galaxies over most of the stellar-mass range, typically accounting together for $\sim70$--$90\%$ of the total population in a given mass bin. In LRG, the partition is even more concentrated, with filaments and knots together contributing $\sim90$--$100\%$ of both the red and blue populations across most of the plotted range. Because RRF and RBF are normalized over environments, these trends should be interpreted as a redistribution of the total red and blue populations across the web, rather than as the probability that a galaxy in a given environment is red or blue.

The BGS panels, shown in Figure \ref{fig:bgs_rrf_rbf_mass}, exhibit that the partition of both red and blue galaxies is dominated by sheets and filaments throughout the observed mass range. In the red population, sheets and filaments typically contribute at the level of $\sim30$--$60\%$ each, while knots and voids usually remain below $\sim10$--$15\%$. The relative balance between sheets and filaments varies with stellar mass and redshift, but no single monotonic crossover is seen in all bins. At intermediate and high masses, filaments often contribute $\sim40$--$65\%$ of the red population, while sheets remain comparably important at $\sim30$--$50\%$. The blue population follows a similarly sheet--filament-dominated partition, with sheets generally contributing $\sim40$--$60\%$ and filaments $\sim30$--$50\%$, whereas voids and knots usually account for only a few percent up to at most $\sim10$--$15\%$. The strong rise of the sheet contribution in the highest-mass part of the lowest-redshift panel, where it approaches unity, should be interpreted with caution, as it is likely amplified by small-number statistics in the extreme high-mass tail. Overall, the BGS results indicate that both quenched and star-forming galaxies are predominantly distributed along the sheet--filament network, with the densest nodes contributing only a modest fraction of the total population.

The LRG sample, shown in Figure \ref{fig:lrg_rrf_rbf_mass}, exhibits a markedly different partition. 
Here the relative fractions are concentrated almost entirely in filaments and knots, while sheets and voids contribute only a very small share at nearly all masses. In the red population, knots typically account for $\sim45$--$70\%$ and filaments for $\sim30$--$55\%$, whereas sheets and voids are generally at the level of only a few percent and are often consistent with zero within the uncertainties. In the $0.6\leq z<0.7$ panel, the knot contribution reaches particularly high values, rising to $\sim70$--$95\%$ at the highest masses, while the filament share correspondingly decreases. By $0.8\leq z<0.9$, however, the two dominant components become more comparable, with knots and filaments each contributing roughly $\sim45$--$60\%$ over much of the mass range. The same overall behaviour is seen for the blue LRG population: although blue LRGs are much less abundant in an absolute sense, their relative distribution is again overwhelmingly split between filaments and knots, with each typically contributing $\sim35$--$60\%$, while sheets and voids remain negligible except in the noisiest low-mass bins. This indicates that even the residual blue component of the LRG sample is largely restricted to the denser structures of the cosmic web. The similarity between the red and blue partitions also suggests that the environmental bias of the tracer itself plays a major role in shaping the relative fractions.

The ELG results, shown in Figure \ref{fig:elg_rrf_rbf_mass}, broadly recover a pattern similar to that seen in BGS. Across most of the well-populated stellar-mass range, sheets and filaments host the largest fractions of both red and blue galaxies, indicating that the ELG population is predominantly distributed along intermediate-density structures rather than concentrated in the densest nodes. For the red population, the relative red fraction (RRF) is generally dominated by sheets and filaments, which typically contribute at the level of roughly \(40\%\)–\(55\%\) and \(35\%\)–\(50\%\), respectively, over most of the better-populated mass bins. By contrast, voids usually contribute only about \(3\%\)–\(8\%\), while knots are generally subdominant at about \(2\%\)–\(10\%\), although they can become slightly larger in some bins at higher redshift. The relative balance between sheets and filaments varies with redshift and, to some extent, with stellar mass: in some panels the sheet contribution is mildly larger, whereas in others the filament contribution becomes comparable or somewhat higher.

A similar partition is found for the blue population. Over most of the well-sampled mass range, sheets and filaments again dominate the relative blue fraction (RBF), typically contributing about \(35\%\)–\(60\%\) and \(30\%\)–\(60\%\), respectively. Voids generally remain at only a few percent, while knots usually contribute at the level of a few percent up to about \(10\%\), and can occasionally rise somewhat above this in individual bins. At the lowest-mass end, both RRF and RBF show sharp excursions, including occasional spikes approaching unity in single bins; these are associated with very small galaxy counts and should therefore not be over-interpreted. Overall, the ELG relative fractions indicate that both red and blue ELGs are primarily associated with the sheet--filament network, while knots provide a smaller, though not always negligible, contribution.

\subsection{Colour distributions across cosmic web environments}
\label{sec:colour_pdf}

\begin{figure}[!htbp]
\centering
\includegraphics[width=0.8\textwidth]{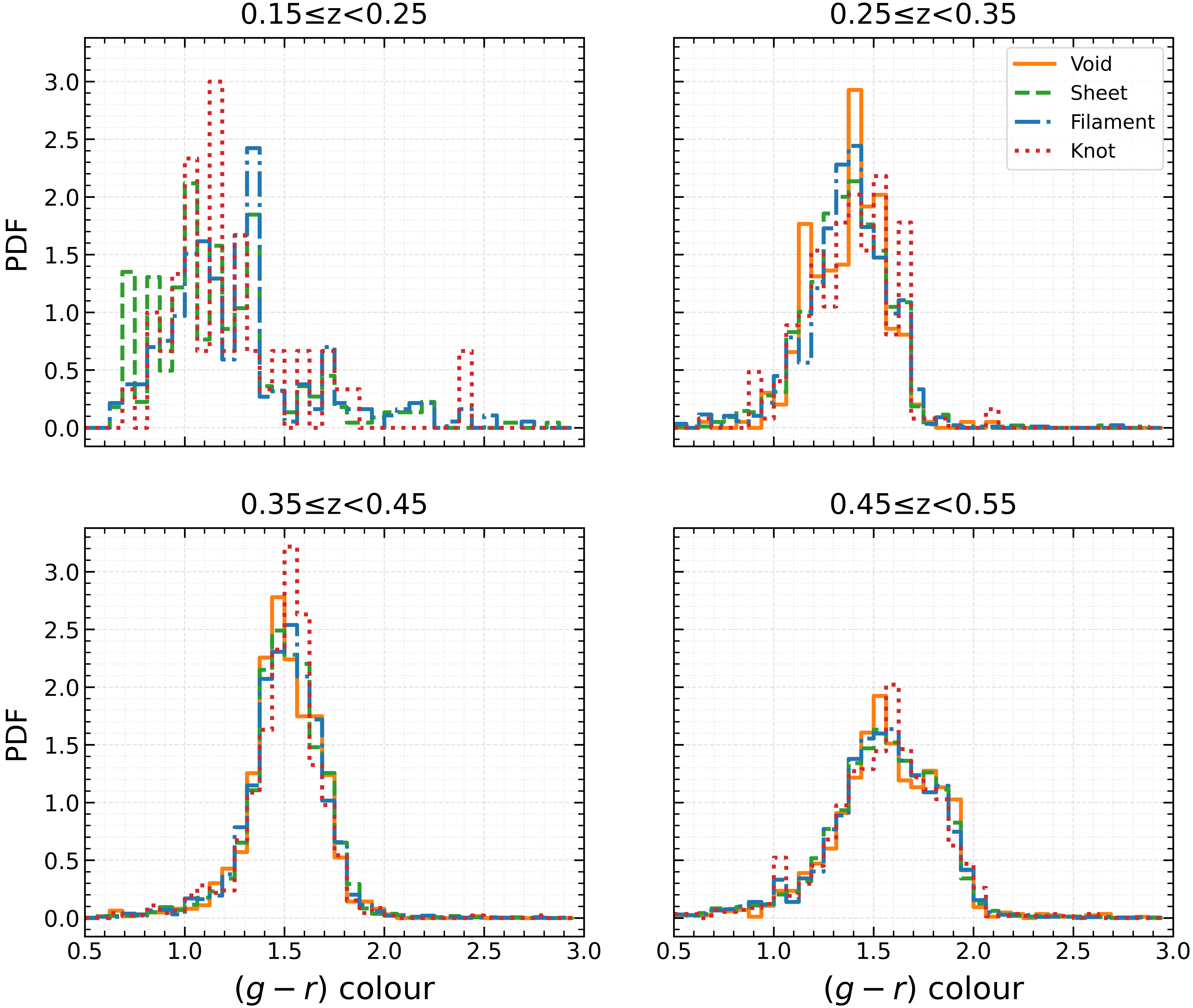}
\caption{Probability distribution functions of observed $(g-r)$ colour across cosmic-web environments for BGS galaxies at $0.15\leq z<0.55$. The curves are normalised to unit area in each environment.}
\label{fig:bgs_colour_pdf}
\end{figure}

\begin{figure}[!htbp]
\centering
\includegraphics[width=0.8\textwidth]{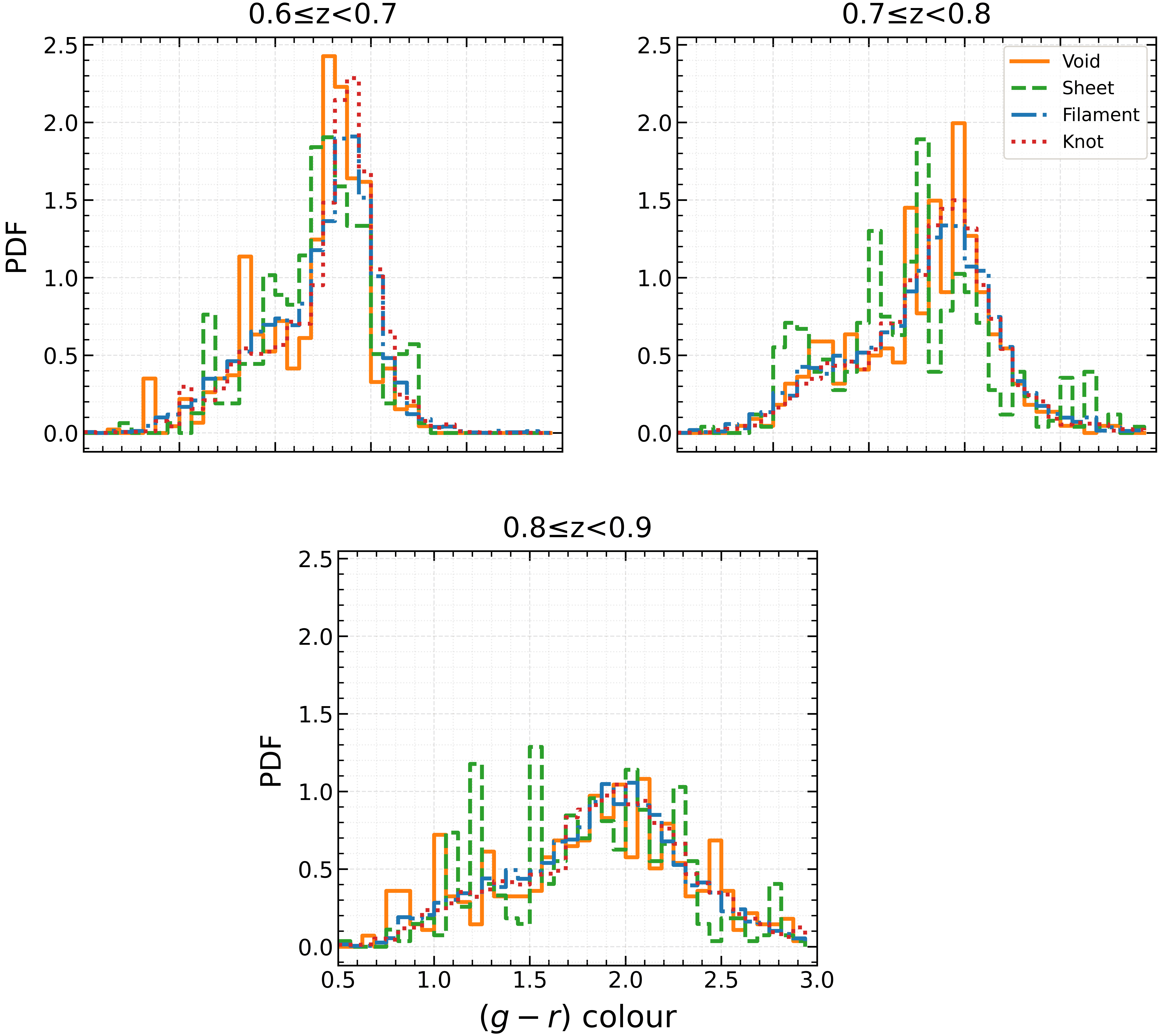}
\caption{Probability distribution functions of observed $(g-r)$ colour across cosmic-web environments for LRG galaxies at $0.6\leq z<0.9$. The curves are normalised to unit area in each environment.}
\label{fig:lrg_colour_pdf}
\end{figure}

\begin{figure}[!htbp]
\centering
\includegraphics[width=1.0\textwidth]{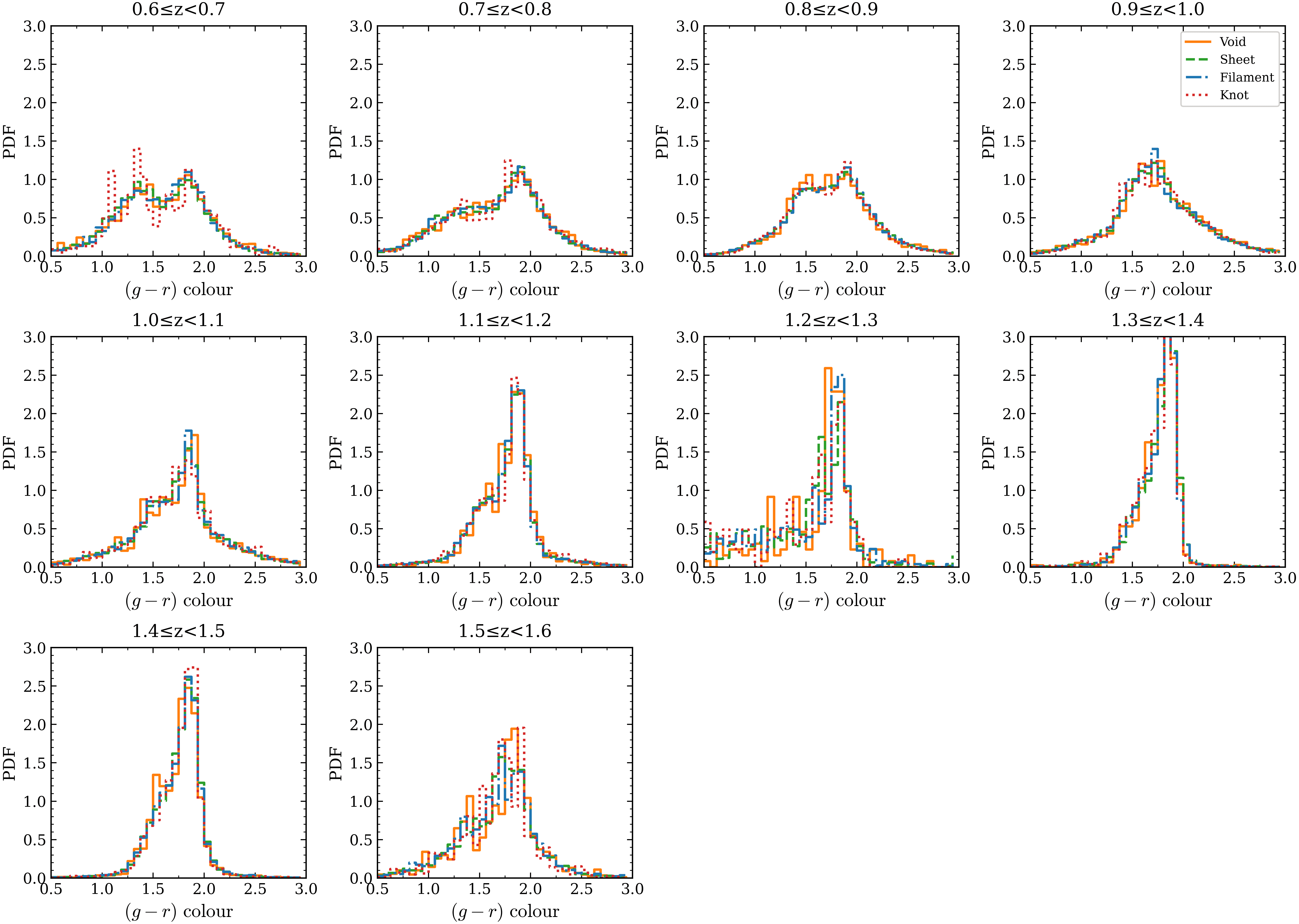}
\caption{Probability distribution functions of observed $(g-r)$ colour across cosmic-web environments for ELG galaxies at $0.6\leq z<1.6$. The curves are normalised to unit area in each environment.}
\label{fig:elg_colour_pdf}
\end{figure}

We now examine the probability distribution functions (PDFs) of observed $(g-r)$ colour across the four cosmic-web environments. Figure~\ref{fig:bgs_colour_pdf} shows the results for BGS, Figure~\ref{fig:lrg_colour_pdf} for LRG, and Figure~\ref{fig:elg_colour_pdf} for ELG. 
The PDFs are normalised to unit area in each environment and redshift bin, highlighting changes in the \emph{shape} of the colour distribution rather than absolute number counts.

The BGS colour distributions, shown in Figure \ref{fig:bgs_colour_pdf}, are broadest and most irregular in the lowest-redshift bin, $0.15\leq z<0.25$, where the mean colours are $1.182\pm0.480$ in voids, $1.181\pm0.375$ in sheets, $1.271\pm0.431$ in filaments, and $1.229\pm0.368$ in knots. These relatively large dispersions are consistent with the visibly noisy PDFs in that panel. At higher redshift, the distributions become smoother and shift systematically to redder colours. In the $0.25\leq z<0.35$ bin, the mean colours lie in the narrow range $1.364$--$1.396$, with standard deviations of $0.194$--$0.282$. In the $0.35\leq z<0.45$ bin, the means increase further to $1.499$--$1.510$, with somewhat smaller dispersions in sheets and filaments ($0.222$ and $0.205$, respectively). In the highest bin, $0.45\leq z<0.55$, the means are $1.562\pm0.280$ in voids, $1.546\pm0.287$ in sheets, $1.549\pm0.299$ in filaments, and $1.533\pm0.324$ in knots. Overall, the BGS PDFs are dominated by a common main peak near $(g-r)\simeq1.4$--$1.6$, and the numerical summaries show that the redshift evolution of the mean colour is stronger than the environment-to-environment variation at fixed redshift. The environmental dependence is therefore modest, appearing mainly through small changes in the width of the distribution and in the strength of the red tail.

The LRG distributions, shown in Figure \ref{fig:lrg_colour_pdf}, are systematically redder than those of BGS and are centered at approximately $(g-r)\sim1.5$--$1.8$, consistent with the intrinsically quiescent nature of the LRG sample. In the $0.6\leq z<0.7$ bin, the mean colours are $1.499\pm0.531$ in voids, $1.545\pm0.355$ in sheets, $1.561\pm0.510$ in filaments, and $1.528\pm0.575$ in knots, indicating broad distributions with only modest separation in the means. In the $0.7\leq z<0.8$ bin, the environmental contrast becomes more visible: sheets are distinctly bluer, with a mean colour of $1.500\pm0.441$, compared with $1.661\pm0.547$ in filaments and $1.720\pm0.503$ in knots, while voids have the reddest mean, $1.784\pm0.600$, albeit with a large dispersion. By $0.8\leq z<0.9$, the distributions remain broad and largely unimodal, with mean colours of $1.746\pm0.552$ in voids, $1.678\pm0.485$ in sheets, $1.785\pm0.455$ in filaments, and $1.818\pm0.474$ in knots. These values support the visual impression that knots and filaments tend to maintain slightly stronger high-colour support than sheets, while the sizeable standard deviations ($\sim0.45$--$0.60$) show that the LRG colour distributions remain broad in all environments. The LRG data therefore indicate a generally red population with mild but detectable environmental modulation, but the claim of a clear bimodal transition in every environment should be stated cautiously.

Figure \ref{fig:elg_colour_pdf} shows the ELG colour distributions are centered at approximately \((g-r)\sim 1.6\)--\(1.8\), placing
them in an intermediate colour regime. In the \(0.6 \leq z < 0.7\) bin, the mean colours are
\(1.597 \pm 0.440\) in voids, \(1.576 \pm 0.424\) in sheets, \(1.601 \pm 0.420\) in filaments, and
\(1.590 \pm 0.429\) in knots, indicating broad distributions with very modest environmental
separation. In the \(0.7 \leq z < 0.8\) bin, the distributions shift systematically toward redder
colours, with mean values of \(1.673 \pm 0.457\), \(1.678 \pm 0.458\), \(1.678 \pm 0.457\), and
\(1.696 \pm 0.466\) in voids, sheets, filaments, and knots, respectively. By \(0.8 \leq z < 0.9\),
the distributions become redder still, with mean colours of \(1.737 \pm 0.406\) in voids,
\(1.745 \pm 0.408\) in sheets, \(1.745 \pm 0.407\) in filaments, and \(1.738 \pm 0.402\) in knots.
The median colours in this interval remain tightly grouped around \((g-r)\approx 1.73\)--\(1.76\),
showing that the environmental differences are small compared with the overall redshift
evolution. The ELG sample therefore exhibits broad colour distributions with mild but
detectable environmental modulation, while the dominant behaviour is a gradual shift toward
redder colours with increasing redshift.

\FloatBarrier

Taken together, the BGS and LRG numerical summaries confirm that the environment dependence in the DESI colour PDFs is expressed mainly through modest shifts in the mean colour and changes in the distribution width, rather than through dramatic changes in the location of the dominant peak. In BGS, the mean colour increases substantially with redshift, from roughly $1.18$--$1.27$ at $0.15\leq z<0.25$ to $1.53$--$1.56$ at $0.45\leq z<0.55$, while the inter-environment differences at fixed redshift remain small. In LRG, the mean colours are systematically redder overall, spanning $\sim1.50$--$1.82$, and the environmental separation is most noticeable in the intermediate and highest redshift bins, where knots and filaments tend to occupy the redder side of the distribution.

\subsection{Comparison with IllustrisTNG predictions}
\label{sec:comparison_tng}

The recent work of \citep{Pandey2025} provides a comprehensive analysis of red and blue galaxy evolution across cosmic web environments using the IllustrisTNG300 simulation, offering a natural benchmark for our DESI DR1 measurements. 
Both studies employ the same methodological framework: a tidal-tensor (T-Web) classification of environments and a mass-dependent Otsu threshold for red/blue separation. 
This common methodology enables a direct comparison between simulation and observations, allowing us to assess the extent to which TNG reproduces the environmental quenching patterns seen in the real universe.

A key finding of \cite{Pandey2025} is that stellar mass is the primary driver of quenching, with environment providing a secondary modulation.
They report that at $z<1$, galaxies with $\log_{10}(M_\ast/M_\odot)>10.5$ are predominantly quenched across all environments, while lower-mass galaxies exhibit stronger environmental dependence.
Our DESI results are in excellent qualitative agreement with this picture. 
For LRGs and ELGs, which sample the high-mass regime, we find red fractions that are consistently elevated across all environments, with environmental differences becoming most apparent at intermediate masses ($\log_{10}(M_\ast/M_\odot)\lesssim10.5$). 
In BGS, the mass-dependent trends are particularly clear: the red fraction rises steeply with stellar mass in all environments, reaching $\sim80\%$ at $\log_{10}(M_\ast/M_\odot)\gtrsim11.2$, while environmental modulation is modest but systematic.

The redshift evolution of red fractions shows notable consistency between TNG and DESI.
\citep{Pandey2025} find that clusters exhibit the earliest rise in red fraction, with an inflection point at $z\sim1.5$, followed by filaments ($z\sim0.7$), sheets, and voids ($z\sim0.5$). 
In our LRG sample ($0.6\le z<0.9$), we observe that knots consistently show the highest red fractions across the full redshift range, with the environmental hierarchy strengthening toward lower redshift.
At $z\simeq0.87$, the knot red fraction reaches $60.5\%\pm1.1\%$, while filaments and sheets are $\sim54\%$; by $z\simeq0.67$, the knot red fraction remains highest at $56.9\%\pm0.5\%$, while sheets drop to $40.5\%\pm3.5\%$.
This progressive divergence mirrors the TNG prediction that environmental quenching becomes increasingly important at late times.

The relative fractions (RRF and RBF) provide another useful point of comparison with the results of \citep{Pandey2025}. They report that in TNG, filaments host the largest share of blue galaxies at all redshifts, while clusters increasingly dominate the red population toward $z=0$. They also note a counter-intuitive trend for high-mass galaxies, namely that clusters show the highest RBF at low redshift for $\log_{10}(M_\ast/M_\odot)>11$. Our DESI measurements support this picture only partially. For the BGS sample, the comparison should be stated more cautiously. After verifying the internal consistency of the BGS fraction calculations using the final mass-dependent Otsu classification, we do not recover the same high-mass cluster-RBF behaviour in DESI. In the lowest BGS redshift bin, $0.15 \le z < 0.25$, the high-mass blue population is dominated by sheets and filaments rather than knots. For $10.5 < \log_{10}(M_\ast/M_\odot) < 12.5$, we obtain $\mathrm{RBF}=0.4646$ in sheets and $\mathrm{RBF}=0.4187$ in filaments, while voids and knots contribute much smaller shares, $\mathrm{RBF}=0.0621$ and $\mathrm{RBF}=0.0546$, respectively. Using the stricter cut $\log_{10}(M_\ast/M_\odot)>11$, the same qualitative result remains: sheets provide the largest contribution, $\mathrm{RBF}=0.5262$, followed by filaments with $\mathrm{RBF}=0.3609$, whereas voids and knots remain minor contributors with $\mathrm{RBF}=0.0484$ and $\mathrm{RBF}=0.0645$. Thus, the DESI BGS sample does not reproduce the specific TNG result that high-mass blue galaxies are most concentrated in the densest environments. Instead, the largest share of such galaxies resides in sheets, with filaments providing the second-largest contribution. A similar behaviour is seen for ELGs, where sheets and filaments again host the largest share of high-mass blue galaxies, with $\mathrm{RBF}\sim0.35$--$0.60$ and $\sim0.30$--$0.50$, respectively, while knots usually remain at the level of only a few per cent up to $\sim0.1$. The only tracer showing partial qualitative agreement with the TNG result is the LRG sample. For $\log_{10}(M_\ast/M_\odot)>11$, the knot contribution is frequently comparable to, or larger than, the filament contribution: at $0.7\leq z<0.8$, knots contribute $\mathrm{RBF}\sim0.43$--$0.69$ compared with $\sim0.30$--$0.58$ for filaments, while at $0.8\leq z<0.9$ the two remain close, with knots at $\sim0.47$--$0.61$ and filaments at $\sim0.41$--$0.50$. Thus, DESI does not show a universal excess of high-mass blue galaxies in the densest environments. Instead, such a tendency appears only in the LRG sample, whereas BGS and ELGs continue to place most high-mass blue galaxies in sheets and filaments. This difference may reflect the distinct selection functions of the DESI tracers relative to the full simulated galaxy population, together with the limited statistical precision in the highest-mass bins.

The ELG results indicate that the dominant observational trend in DESI DR1 is a sheet--filament partition rather than a filament--knot dominance \citep{Pandey2025}. This conclusion is supported consistently by both sets of diagnostics. In the RF/BF analysis, the red fraction increases with stellar mass and evolves strongly with redshift, but the environmental ordering is comparatively weak and not fixed across all bins; thus, although denser environments can show moderately enhanced red fractions, knots do not emerge as the dominant environment in a systematic way. In the complementary RRF/RBF analysis, which measures the share of the total red or blue population contributed by each environment within a given mass--redshift bin, sheets and filaments host the largest fractions over most of the well-populated range, while voids remain minor contributors and knots are generally secondary. This behaviour is physically expected for an observational ELG sample, because sheets and filaments contain most of the galaxies overall and therefore naturally contribute the largest relative fractions, even if knots correspond to the densest local environments. Quantitatively, over the better-sampled mass bins, the red relative fraction is typically split between sheets and filaments at roughly \(40\%\)–\(55\%\) and \(35\%\)–\(50\%\), respectively, while the blue relative fraction is also mainly shared between sheets and filaments, typically at the level of about \(35\%\)–\(60\%\) and \(30\%\)–\(55\%\); by contrast, voids and knots are usually much smaller contributors. At the lowest-mass end, both RRF and RBF show occasional sharp excursions, including spikes toward unity in single bins, but these coincide with very small red or blue counts and are therefore not interpreted as robust astrophysical trends. Overall, unlike the stronger filament--knot dominance reported for IllustrisTNG by \citep{Pandey2025}, the DESI DR1 ELG data support a picture in which both the RF/BF and RRF/RBF diagnostics are dominated by the sheet--filament network, with knots providing only a smaller, though non-zero, contribution.

The evolution of colour distributions in TNG shows a transition from unimodal to bimodal by $z=2$, with clusters developing the strongest red sequence \citep{Pandey2025}. 
In DESI, we observe the same qualitative behaviour: at the highest redshifts probed by ELGs ($z\sim1.5$), the colour distributions are broad and show only weak bimodality, while at lower redshifts (LRG at $z\sim0.6$, BGS at $z\sim0.3$), clear bimodality is present in all environments, with knots exhibiting the most pronounced red peak.
The median colour evolution in DESI also mirrors TNG: clusters show the highest median $(g-r)$ at $z<1$, followed by filaments, sheets, and voids.
For massive galaxies, we find that median colour varies little across environments, consistent with the TNG conclusion that colour evolution in this regime is primarily mass-driven.

One area where DESI and TNG show subtle differences is in the amplitude of environmental modulation.
In TNG, \citep{Pandey2025} report red fractions in clusters at $z=0$ of $\sim75\%$, compared to $\sim45\%$ in filaments, $\sim34\%$ in sheets, and $\sim30\%$ in voids. 
Our LRG measurements at $z\simeq0.67$ show a cluster red fraction of $\sim57\%$, filaments $\sim50\%$, sheets $\sim40\%$, and voids $\sim51\%$. While the environmental ordering is similar, the dynamic range in DESI appears somewhat compressed, particularly for voids. 
This may reflect observational selection effects: DESI LRGs are massive, quenched galaxies by design, so even galaxies in underdense regions are pre-selected to be red. 
Additionally, the T-Web reconstruction in observational data is affected by redshift-space distortions and survey boundaries, which could smooth environmental contrasts relative to the simulation.

The TNG analysis emphasizes the role of AGN feedback in quenching massive galaxies across all environments, while environmental processes (ram-pressure stripping, strangulation, mergers) dominate at lower masses \citep{Donnari2021,Pandey2025}. 
Our DESI results are fully consistent with this two-channel quenching paradigm. 
The steep mass dependence of the red fraction in all environments points to a mass-driven process that operates universally, while the residual environmental ordering at fixed mass---particularly evident at $\log_{10}(M_\ast/M_\odot)\lesssim10.5$---requires additional quenching mechanisms that are more efficient in dense regions.
The consistency between TNG predictions and DESI observations suggests that the physical processes implemented in the simulation (AGN feedback, gas stripping, mergers) capture the essential ingredients governing galaxy transformation across the cosmic web.

In summary, our DESI DR1 analysis provides strong observational support for the environmental quenching framework developed from IllustrisTNG. 
The qualitative agreement extends across multiple metrics: redshift evolution of red fractions, mass dependence of environmental contrast, partitioning of populations across the web, and colour evolution. 
Quantitative differences remain, particularly in the amplitude of environmental effects, but these are plausibly attributable to selection effects and observational systematics rather than fundamental shortcomings of the simulation. 
Future work with DESI data releases, incorporating realistic mock catalogs and forward-modeling of selection effects, will enable even more precise comparisons and help refine the physical models of galaxy quenching.

\section{Conclusions}\label{sec:conclusions}

We have conducted a systematic, multi-tracer analysis of galaxy quenching in the cosmic web using DESI DR1, combining a tidal-tensor (T-Web) reconstruction of large-scale environments with a mass-dependent Otsu classification of galaxies into red (quenched) and blue (star-forming) populations. Our main findings can be summarised as follows.

Stellar mass is the primary driver of quenching across all tracers, consistent with the established paradigm that mass quenching dominates over environmental effects at high masses \citep{Peng2010}. 
In every environment and redshift bin, the red fraction increases monotonically with stellar mass, with the blue fraction decreasing accordingly.
For BGS, the global red fraction rises from approximately 0.50 at $0.15\le z<0.25$ to 0.72 at $0.35\le z<0.45$, reaching approximately 0.80 at the highest masses ($\log_{10}(M_\ast/M_\odot)\gtrsim11.2$). 
This universal trend confirms that mass quenching through internal and halo-related processes is the dominant mechanism regulating star formation.

At fixed stellar mass and redshift, the cosmic-web environment provides a systematic secondary modulation. 
Knots consistently exhibit the highest red fractions, followed by filaments and sheets, with voids showing the lowest.
For LRG at $z\simeq0.87$, the red fraction reaches $60.5\%$ in knots compared to approximately $54\%$ in filaments and sheets; by $z\simeq0.67$, the knot red fraction remains highest at $56.9\%$, while sheets drop to $40.5\%$. 
This environmental hierarchy is consistent with enhanced processing in dense regions through mechanisms such as ram-pressure stripping, strangulation, and mergers, and weaker processing in underdense regions. 
The environmental contrast is strongest at intermediate stellar masses ($\log_{10}(M_\ast/M_\odot)\lesssim10.5$), where galaxies are more susceptible to external effects, and weakens at the highest masses where mass quenching dominates.

The relative fractions RRF and RBF reveal how red and blue populations are partitioned across the cosmic web. 
Filaments and sheets host the largest shares of both populations, reflecting their dominant volume fraction in the survey. 
Knots, despite being efficient quenching sites, contribute only a small fraction of the total red population (typically $5$--$10\%$) because they are volumetrically rare. 
The knot contribution to the red population increases with stellar mass, indicating that the most massive quenched galaxies preferentially reside in the densest environments. 
Colour PDFs reinforce this picture, showing clear bimodality at low redshift in all environments, with dense regions exhibiting an enhanced red sequence. 
The gradual redward drift of the blue cloud and the strengthening of the red sequence toward lower redshift are consistent with ageing stellar populations and the build-up of the quenched population, with knots leading this evolution.

These results demonstrate that while stellar mass sets the primary quenching scale, the cosmic-web environment provides a systematic and physically meaningful secondary modulation. 
The consistency of trends across three independent tracers with their different bias, redshift ranges, and selection functions validates the robustness of our findings and underscores the importance of multi-tracer analyses for environmental inference.

The framework presented here opens several avenues for future work.
A quantitative comparison with hydrodynamic simulations, applying the same measurement pipeline to mock catalogs from IllustrisTNG, will enable a direct test of the environmental quenching predictions from \citep{Pandey2025} and move beyond the qualitative agreement reported here.
Systematic exploration of T-Web parameters (smoothing scale, grid resolution, eigenvalue threshold) and survey-window effects will quantify the sensitivity of our environmental metrics.
Extending the analysis to the full ELG redshift range in DR1 and to future DESI data releases will constrain the onset and evolution of environmental quenching over a wider cosmic baseline.
Overall, this work establishes a reproducible and extensible methodology for precision studies of galaxy evolution in the cosmic web with DESI, and provides the first observational validation of key predictions from state-of-the-art hydrodynamic simulations using the largest spectroscopic sample ever assembled.
Joint modelling approaches that simultaneously describe the mass and environment dependence of quenching can disentangle their relative contributions, while cross-validation with alternative cosmic-web classifiers will establish the robustness of our conclusions to the specific definition of environment.

\section{Data Availability}

The data underlying this article will be shared by the corresponding author upon reasonable request.

\section*{Acknowledgements}
HIU and MA contributed equally to this work. HIU and MA gratefully acknowledge the financial support of the SECIHTI PhD scholarships.
This work was partially supported by SECIHTI M\'exico under grants CBF-2025-G-1720 and CBF-2025-G-176. Also  by the grant I0101/131/07 C-234/07 of the Instituto Avanzado de Cosmolog\'ia (IAC) collaboration (http://www.iac.edu.mx/), and for the computing time granted by LANCAD and SECIHTI in the Supercomputer Hybrid Cluster ``Xiuhcoatl" at GENERAL COORDINATION OF INFORMATION AND COMMUNICATIONS TECHNOLOGIES (CGSTIC) of CINVESTAV.
URL: http://clusterhibrido.cinvestav.mx/

\appendix
\section{T-Web supplementary material}
\label{app:tweb_tables}

This appendix provides the complete volume fractions for each tracer as a function of smoothing scale, along with four-panel decompositions and schematic illustrations of the T-Web classification.

\subsection{BGS: volume fractions and four-panel decomposition}

\begin{table}[H]
\centering
\begin{tabular}{|c|c|c|c|c|}
\hline
$R_s$ [Mpc] & Voids [\%] & Sheets [\%] & Filaments [\%] & Knots [\%] \\
\hline
6  & 15.6 & 46.0 & 38.0 & 5.2 \\
7  & 15.5 & 46.0 & 38.2 & 5.3 \\
9  & 15.4 & 45.8 & 38.3 & 5.4 \\
10 & 15.3 & 45.6 & 38.5 & 5.5 \\
11 & 15.2 & 45.5 & 38.5 & 5.6 \\
\hline
\end{tabular}
\caption{Complete T-Web environment volume fractions for the BGS reconstruction as a function of smoothing scale.}
\label{tab:tweb_fractions_bgs_app}
\end{table}

\begin{figure}[H]
\centering
\includegraphics[width=0.67\textwidth]{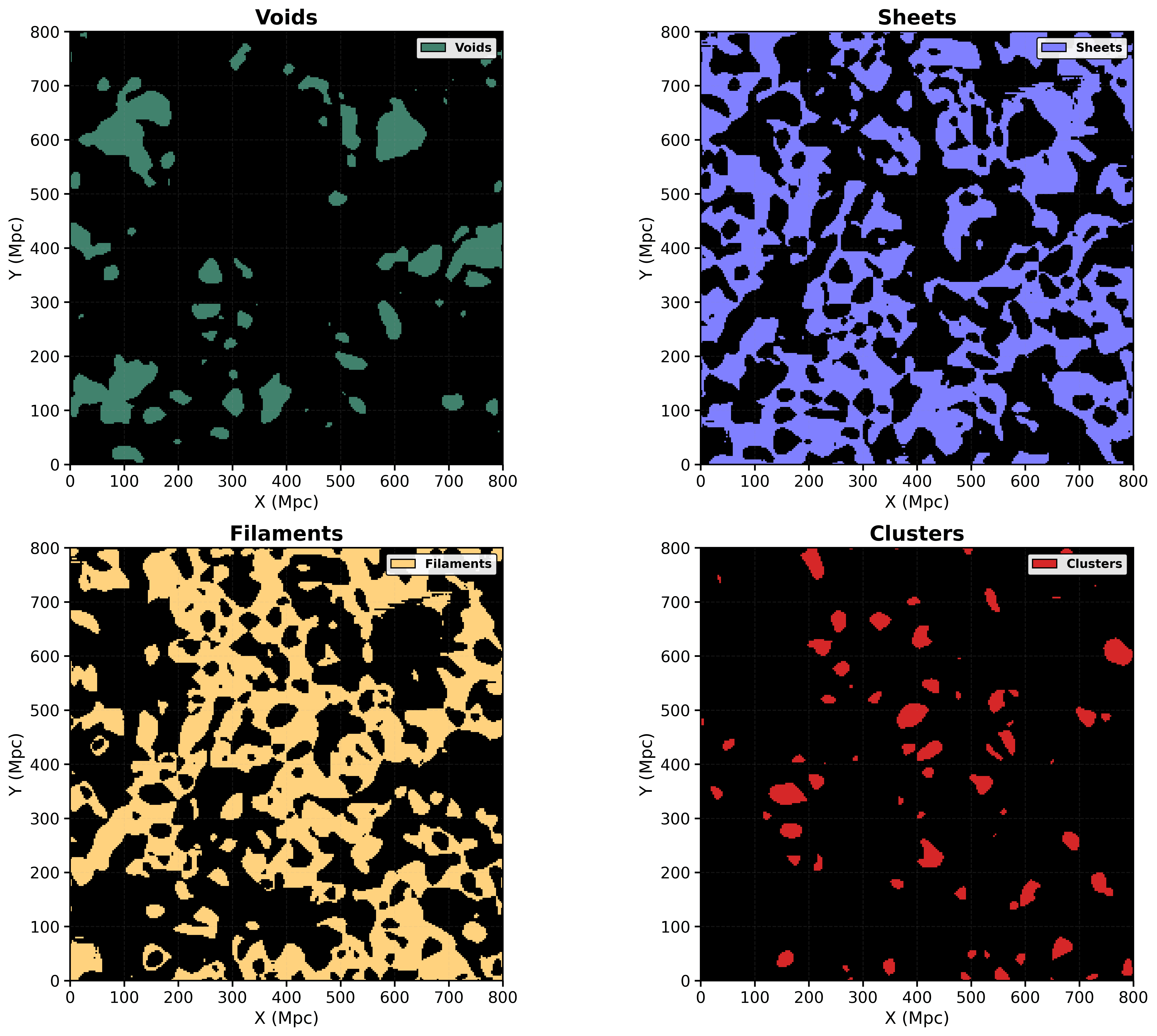}
\caption{Four-panel representation of the BGS T-Web classification, highlighting the spatial regions assigned to voids, sheets, filaments, and knots.}
\label{fig:tweb_bgs_panels_app}
\end{figure}

\subsection{LRG: volume fractions and four-panel decomposition}

\begin{table}[H]
\centering
\begin{tabular}{|c|c|c|c|c|}
\hline
$R_s$ [Mpc] & Voids [\%] & Sheets [\%] & Filaments [\%] & Knots [\%] \\
\hline
7  & 10.0 & 46.9 & 37.8 & 5.3 \\
9  & 9.7 & 46.2 & 38.5 & 5.6 \\
11 & 9.5 & 45.6 & 39.0 & 5.9 \\
12 & 9.3 & 45.3 & 39.3 & 6.0 \\
14 & 9.1 & 44.8 & 39.7 & 6.4 \\
\hline
\end{tabular}
\caption{Complete T-Web environment volume fractions for the LRG reconstruction as a function of smoothing scale.}
\label{tab:tweb_fractions_lrg_app}
\end{table}

\subsection{ELG: volume fractions and schematic illustration}

\begin{table}[H]
\centering
\begin{tabular}{|c|c|c|c|c|}
\hline
$R_s$ [Mpc] & Voids [\%] & Sheets [\%] & Filaments [\%] & Knots [\%] \\
\hline
2  & 6.8  & 46.7 & 40.4 & 6.1 \\
5  & 5.7  & 48.1 & 39.6 & 6.6 \\
10 & 7.4  & 47.1 & 39.2 & 6.3 \\
15 & 10.4 & 45.9 & 37.8 & 5.8 \\
20 & 12.5 & 46.5 & 36.6 & 4.4 \\
\hline
\end{tabular}
\caption{Complete T-Web environment volume fractions for the ELG reconstruction as a function of smoothing scale.}
\label{tab:tweb_fractions_elg_app}
\end{table}

\begin{figure}[H]
\centering
\includegraphics[width=0.67\textwidth]{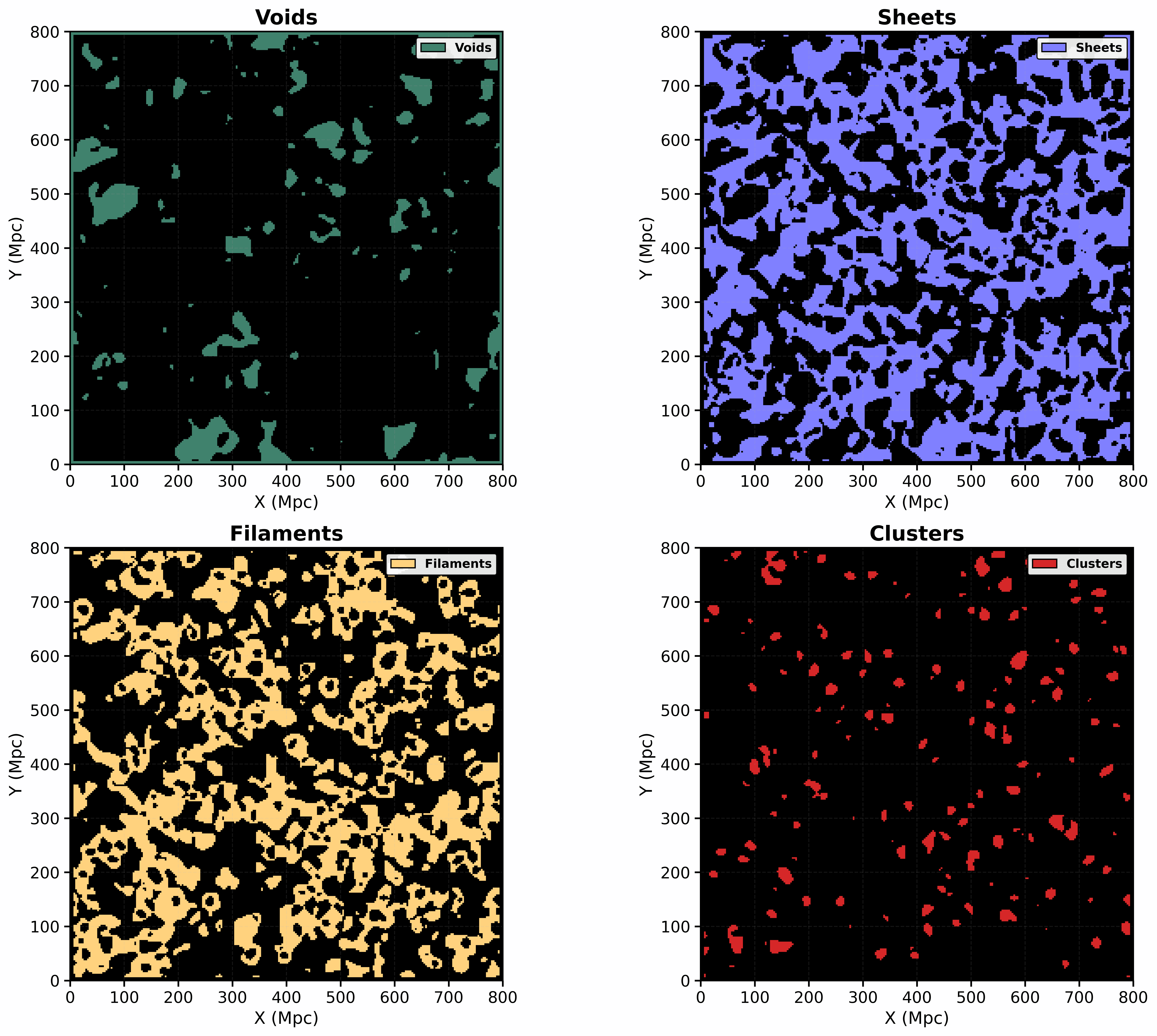}
\caption{Four-panel representation of the LRG T-Web classification.}
\label{fig:tweb_lrg_panels_app}
\end{figure}

\begin{figure}[H]
\centering
\includegraphics[width=0.67\textwidth]{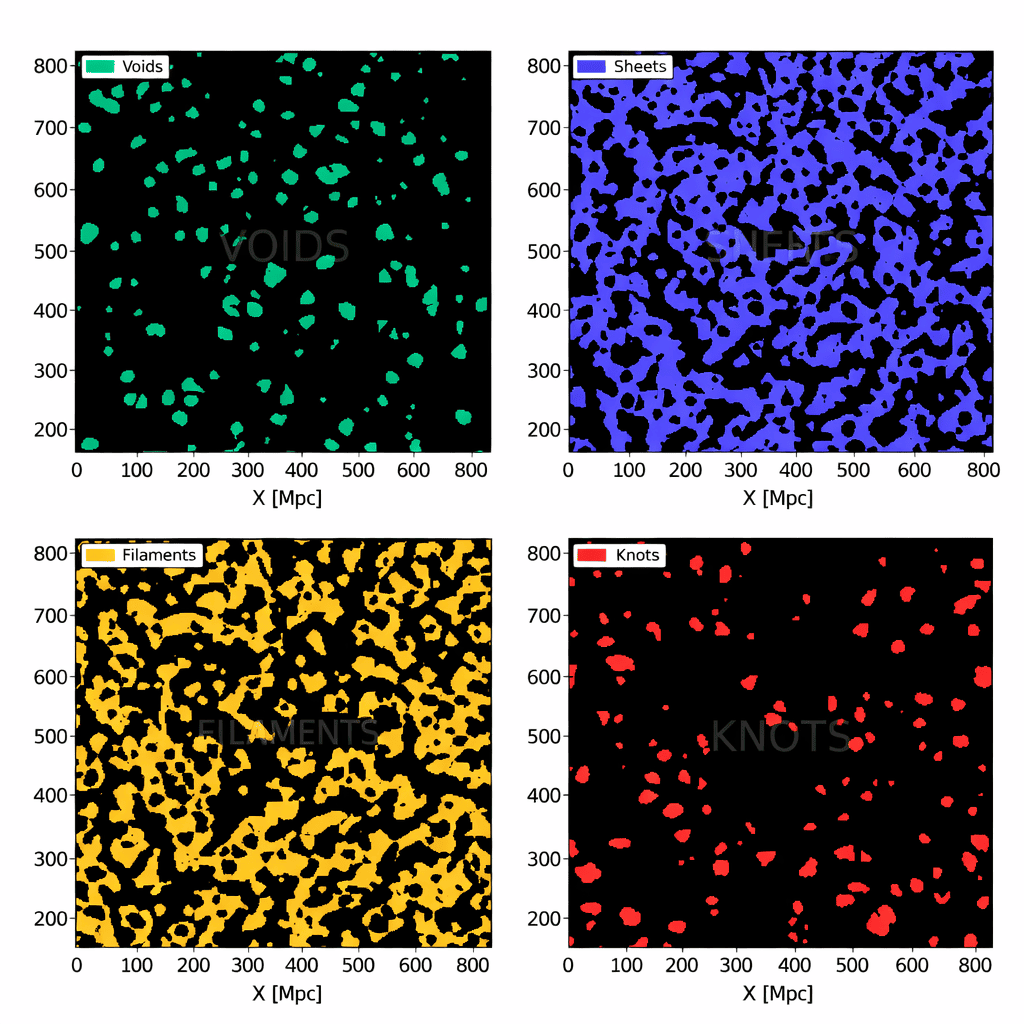}
\caption{Four-panel representation of the ELG T-Web classification.}
\label{fig:tweb_schematic_app}
\end{figure}

\section{Otsu number counts}
\label{app:otsu_counts}

This appendix provides the complete number counts of red and blue galaxies for each DESI tracer, as determined by the mass-dependent Otsu classification.

After applying quality cuts, the total number of galaxies used in this study for each tracer is as follows: \textbf{BGS:} 1,399,434 galaxies. Of these, 1,293,399 lie within the analysis redshift range $0.15\le z<0.55$. \textbf{LRG:} 673,995 galaxies. Of these, 417,980 lie within the analysis redshift range $0.6\le z<0.9$. \textbf{ELG:} 862,541 galaxies within the analysis redshift range $0.6\le z\le1.6$.
In total, the combined DESI DR1 sample used in this study comprises 2,573,920 galaxies after quality cuts.

Tables~\ref{tab:BGS_counts_app}–\ref{tab:ELG_counts_app} list the counts for the full sample and for the high-mass subsample ($10.5 < \log_{10}(M_\star/M_\odot) < 12.5$) in each redshift bin.

\begin{table}[H]
\centering
\begin{tabular}{l|ccc|ccc}
\hline
\multicolumn{1}{c|}{} & \multicolumn{3}{c|}{All} & \multicolumn{3}{c}{$10.5 < \log_{10}(M_\star/M_\odot) < 12.5$} \\
\hline
Redshift & Total & Red & Blue & Total & Red & Blue \\
\hline
0.15--0.25 & 39203  & 19443  & 19760  & 37849  & 19161  & 18688 \\
0.25--0.35 & 219160 & 135411 & 83749  & 218718 & 135273 & 83445 \\
0.35--0.45 & 667894 & 480623 & 187271 & 667696 & 480614 & 187082 \\
0.45--0.55 & 367142 & 255722 & 111420 & 366617 & 255604 & 111013 \\
\hline
\end{tabular}
\caption{BGS red/blue counts.}
\label{tab:BGS_counts_app}
\end{table}

\begin{table}[H]
\centering
\begin{tabular}{l|ccc|ccc}
\hline
\multicolumn{1}{c|}{} & \multicolumn{3}{c|}{All} & \multicolumn{3}{c}{$10.5 < \log_{10}(M_\star/M_\odot) < 12.5$} \\
\hline
Redshift & Total & Red & Blue & Total & Red & Blue \\
\hline
0.6--0.7 & 62144  & 30519  & 31625  & 61645  & 30406  & 31239 \\
0.7--0.8 & 77840  & 45561  & 32279  & 77411  & 45473  & 31938 \\
0.8--0.9 & 277996 & 143638 & 134358 & 277427 & 143520 & 133907 \\
\hline
\end{tabular}
\caption{LRG red/blue counts.}
\label{tab:LRG_counts_app}
\end{table}

\begin{table}[H]
\centering
\begin{tabular}{l|ccc|ccc}
\hline
\multicolumn{1}{c|}{} & \multicolumn{3}{c|}{All} & \multicolumn{3}{c}{$10.5 < \log_{10}(M_\star/M_\odot) < 12.5$} \\
\hline
Redshift & Total & Red & Blue & Total & Red & Blue \\
\hline
0.6--0.7 & 61273  & 28462  & 32811  & 60971  & 28457  & 32514 \\
0.7--0.8 & 76309  & 43307  & 33002  & 76158  & 43306  & 32852 \\
0.8--0.9 & 273025 & 157736 & 115289 & 272801 & 157722 & 115079 \\
0.9--1.0 & 100656 & 55595  & 45061  & 100508 & 55589  & 44919  \\
1.0--1.1 & 57089  & 35002  & 22087  & 56980  & 34998  & 21982  \\
1.1--1.2 & 73505  & 49735  & 23770  & 73389  & 49732  & 23657  \\
1.2--1.3 & 1887   & 1084   & 803    & 1854   & 1083   & 771    \\
1.3--1.4 & 66585  & 48105  & 18480  & 66523  & 48099  & 18424  \\
1.4--1.5 & 132790 & 91301  & 41489  & 132670 & 91289  & 41381  \\
1.5--1.6 & 19422  & 11310  & 8112   & 19295  & 11281  & 8014   \\
\hline
\end{tabular}
\caption{ELG red/blue counts.}
\label{tab:ELG_counts_app}
\end{table}

\bibliographystyle{JHEP}
\bibliography{biblio}

\end{document}